\documentstyle[12pt,epsfig,rotating,a4]{article}
\begin{document}

\begin{titlepage}
\bigskip
\begin{flushright}
   {24.11.2002}
\end{flushright}
\bigskip \bigskip
\begin{center}{\LARGE\bf Higgs Statistics for Pedestrians}
\end{center}
\bigskip
\begin{center}{Eilam Gross and Amit Klier}
\end{center}
\begin{center}
      {Department of  Particle Physics \\
Weizmann Institute of Science, Rehovot 76100,
Israel}\\
     \textit{email: eilam.gross@weizmann.ac.il}
\end{center}


\baselineskip 12pt
\parskip 12 true pt
\setlength {\textheight} {23 true cm} \setlength {\textwidth} {17
true cm} \setlength {\oddsidemargin} {0 mm} \setlength
{\evensidemargin} {0 mm} \setlength {\topmargin} {-5 mm}
\setlength {\headheight} {15 pt} \setlength {\headsep} {30 pt}
\textfloatsep 10 mm
\renewcommand{\descriptionlabel}[1]{\hspace\labelsep
                                  \normalfont #1}

 \vskip 1.cm\noindent

\newcommand{\sfer}{\mbox{$\tilde f$}}
\newcommand{\sq}{\mbox{$\tilde q$}}
\newcommand{\sqbar}{\mbox{$\tilde \bar q$}}
\newcommand{\smuon}{\mbox{$\tilde \mu$}}
\newcommand{\sel}{\mbox{$\tilde e$}}
\newcommand{\stau}{\mbox{$\tilde \tau$}}
\newcommand{\msmuon}{\mbox{$m_{\tilde \mu}$}}
\newcommand{\msel}{\mbox{$m_{\tilde e}$}}
\newcommand{\mstau}{\mbox{$m_{\tilde \tau}$}}
\newcommand{\sfl}{\mbox{$\tilde f_L$}}
\newcommand{\sfr}{\mbox{$\tilde f_R$}}
\newcommand{\slep}{\mbox{$\tilde l$}}
\newcommand{\sll}{\mbox{$\tilde l_L$}}
\newcommand{\slr}{\mbox{$\tilde l_R$}}
\newcommand{\sn}{\mbox{$\tilde \nu$}}
\newcommand{\stp}{\mbox{$\tilde t_1$}}
\newcommand{\st}{\mbox{$\tilde t$}}
\newcommand{\sbotp}{\mbox{$\tilde b_1$}}
\newcommand{\DM}{\mbox{$\Delta M$}}
\newcommand{\mst}{\mbox{$m_{\tilde t}$}}
\newcommand{\msq}{\mbox{$m_{\tilde q}$}}
\newcommand{\msn}{\mbox{$m_{\tilde \nu}$}}
\newcommand{\msf}{\mbox{$m_{\tilde f}$}}
\newcommand{\mslr}{\mbox{$m_{\tilde  l_R}$}}
\newcommand{\msll}{\mbox{$m_{\tilde  l_L}$}}
\newcommand{\msl}{\mbox{$m_{\tilde  l}$}}
\newcommand{\msb}{\mbox{$m_{\tilde b_1}$}}
\newcommand{\neut}{\mbox{$\tilde \chi_1^0$}}
\newcommand{\neutp}{\mbox{$\tilde \chi_2^0$}}
\newcommand{\charg}{\mbox{$\tilde \chi^+_1$}}
\newcommand{\mneut}{\mbox{$m_{\tilde \chi^0_1}$}}
\newcommand{\mneutp}{\mbox{$m_{\tilde \chi^0_2}$}}
\newcommand{\mcharg}{\mbox{$m_{\tilde \chi^+_1}$}}
\newcommand{\grav}{\mbox{$\tilde G$}}
\newcommand{\rp}{\mbox{$R_p$}}
\newcommand{\rpv}{\mbox{$R_pV$}}

\newcommand{\sqrtsp}{\mbox{$\sqrt{s'}$}}
\newcommand{\sqrts}{\mbox{$\sqrt{s}$}}
\newcommand{\epm} {\mbox{$\mathrm{e}^+ \mathrm{e}^-$}}
\newcommand{\mpm} {\mbox{$\mu^+ \mu^-$}}
\newcommand{\nprode}{\mbox{$N^{\mathrm{e}^+ \mathrm{e}^-}_{prod}$}}
\newcommand{\nprodm}{\mbox{$N^{\mu^+ \mu^-}_{prod}$}}
\newcommand{\nexpe}{\mbox{$N_{exp}^{\mathrm{e}^+ \mathrm{e}^-}$}}
\newcommand{\nexpm}{\mbox{$N_{exp}^{\mu^+ \mu^-}$}}
\newcommand{\nexpt}{\mbox{$N_{exp}^{total}$}}
\newcommand{\Zo}{\mbox{$\mathrm{Z}^{0}$}}
\newcommand{\bZo}{{\bf \mbox{$\mathrm{Z}^{0}$}}}
\newcommand{\Zs}{\mbox{$\mathrm{Z}^{0*}$}}
\newcommand{\ho}{\mbox{$\mathrm{h}^{0}$}}
\newcommand{\Ho}{\mbox{$\mathrm{H}^{0}$}}
\newcommand{\Ao}{\mbox{$\mathrm{A}^{0}$}}
\newcommand{\Wpm}{\mbox{$\mathrm{W}^{\pm}$}}
\newcommand{\Hpm}{\mbox{$\mathrm{H}^{\pm}$}}
\newcommand{\WW}{\mbox{$\mathrm{W}^{+}\mathrm{W}^{-}$}}
\newcommand{\ZZ}{\mbox{$\mathrm{Z}^{0}\mathrm{Z}^{0}$}}
\newcommand{\Zg}{\mbox{$\mathrm{Z}^{0}/ \mathrm{\gamma}$}}
\newcommand{\ZZs}{\mbox{$\mathrm{Z}^{0}\mathrm{Z}^{0*}$}}
\newcommand{\Zbb}{\mbox{$\mathrm{Z}^{0}\mathrm{b}\mathrm{\bar b}$}}
\newcommand{\koko}{\mbox{${\tilde{\chi_1}^0}{{\tilde{\chi_1}^0}}$}}
\newcommand{\bho}{\mbox{$\boldmath{\mathrm{H}^{0}}$}}
\newcommand{\ee}{\mbox{$\mathrm{e}^{+}\mathrm{e}^{-}$}}
\newcommand{\bee}{\mbox{$\boldmath {\mathrm{e}^{+}\mathrm{e}^{-}} $}}
\newcommand{\bbbar}{\mbox{$\mathrm{b}\mathrm{\bar b}$}}
\newcommand{\ttbar}{\mbox{$\mathrm{t}\mathrm{\bar t}$}}
\newcommand{\mm}{\mbox{$\mu^{+}\mu^{-}$}}
\newcommand{\bmm}{\mbox{$\boldmath {\mu^{+}\mu^{-}} $}}
\newcommand{\nn}{\mbox{$\nu \bar{\nu}$}}
\newcommand{\bnn}{\mbox{$\boldmath {\nu \bar{\nu}} $}}
\newcommand{\qq}{\mbox{$\mathrm{q} \bar{\mathrm{q}}$}}
\newcommand{\ff}{\mbox{$\mathrm{f} \bar{\mathrm{f}}$}}
\newcommand{\fb}{\mbox{$\mathrm{fb^{-1}}$}}
\newcommand{\bqq}{\mbox{$\boldmath {\mathrm{q} \bar{\mathrm{q}}} $}}
\newcommand{\ra}{\mbox{$\rightarrow$}}

\newcommand{\erh}{\mbox{$\mathrm{e}^+\mathrm{e}^-\rightarrow\mathrm{hadrons}$}}
\newcommand{\tptm}{\mbox{$\tau^{+}\tau^{-}$}}
\newcommand{\tpm}{\mbox{$\tau^{\pm}$}}

\newcommand{\gamgam}{\mbox{$\gamma\gamma$}}
\newcommand{\uu}{\mbox{$\mathrm{u} \bar{\mathrm{u}}$}}
\newcommand{\dd}{\mbox{$\mathrm{d} \bar{\mathrm{d}}$}}
\newcommand{\bb}{\mbox{$\mathrm{b} \bar{\mathrm{b}}$}}
\newcommand{\cc}{\mbox{$\mathrm{c} \bar{\mathrm{c}}$}}
\newcommand{\nunu}{\mbox{$\nu \bar{\nu}$}}
\newcommand{\mZ}{\mbox{$m_{\mathrm{Z}^{0}}$}}
\newcommand{\mH}{\mbox{$m_{\mathrm{H}^{0}}$}}
\newcommand{\mh}{\mbox{$m_{\mathrm{h}^{0}}$}}
\newcommand{\mA}{\mbox{$m_{\mathrm{A}^{0}}$}}
\newcommand{\mHpm}{\mbox{$m_{\mathrm{H}^{\pm}}$}}
\newcommand{\mW}{\mbox{$m_{\mathrm{W}^{\pm}}$}}
\newcommand{\mtop}{\mbox{$m_{\mathrm{t}}$}}
\newcommand{\mb}{\mbox{$m_{\mathrm{b}}$}}
\newcommand{\lpm}{\mbox{$\ell ^+ \ell^-$}}
\newcommand{\G}{\mbox{$\mathrm{GeV}$}}
\newcommand{\Gc}{\mbox{$\mathrm{GeV/c}$}}
\newcommand{\Gcs}{\mbox{$\mathrm{GeV/c^2}$}}
\newcommand{\epsnn}{\mbox{$\epsilon^{\nu\bar{\nu}}$(\%)}}
\newcommand{\Nnn}{\mbox{$N^{\nu \bar{\nu}}_{exp}$}}
\newcommand{\epsll}{\mbox{$\epsilon^{\ell^{+}\ell^{-}}$(\%)}}
\newcommand{\Nll}{\mbox{$N^{\ell^+\ell^-}_{exp}$}}
\newcommand{\Nexp}{\mbox{$N^{total}_{exp}$}}
\newcommand{\kl}{\mbox{$\mathrm{K_{L}}$}}
\newcommand{\dedx}{\mbox{d$E$/d$x$}}
\newcommand{\etal}{\mbox{$et$ $al.$}}
\newcommand{\ie}{\mbox{$i.e.$}}
\newcommand{\sba}{\mbox{$\sin ^2 (\beta -\alpha)$}}
\newcommand{\cba}{\mbox{$\cos ^2 (\beta -\alpha)$}}
\newcommand{\tanb}{\mbox{$\tan \beta$}}
\newcommand{\PhysLett}  {Phys.~Lett.}
\newcommand{\PhysRep}   {Phys.~Rep.}
\newcommand{\PhysRev}   {Phys.~Rev.}
\newcommand{\NPhys}  {Nucl.~Phys.}
\newcommand{\CPC} {Comp.~Phys.\ Comm.}
\newcommand{\ZPhys}  {Z.~Phys.}
\newcommand{\IEEENS} {IEEE Trans.\ Nucl.~Sci.}

\begin{abstract}
We review the results of the Standard Model Higgs boson search at
LEP.  An emphasis is put on revealing the details behind the
statistical procedure developed by the LEP Higgs working group.
The procedure is explained using a toy model which allows the
reader to estimate the significance of the experimental
observation which led at the time to a scientific debate on
whether LEP has observed a 115 GeV Higgs boson.

\end{abstract}

\end{titlepage}

\section{Introduction}

On 3 November 2000 in a seminar at CERN the LEP Higgs working
group presented preliminary results of an analysis indicating a
possible 2.9$\sigma$ observation of a 115 GeV Higgs boson
\cite{bib-pik}. Based on this analysis the four LEP collaborations
requested the continuation of LEP  to collect more data at $\sqrt
s = 208$ GeV. However, the arguments presented by the LEP
collaborations did not convince the LEP management  and
 in retrospect, it turned out that  the LEP accelerator turn-off date of 2 November 2000 ended its
 eleven years of forefront research.

\begin{figure}
  \centering
  \epsfig{file=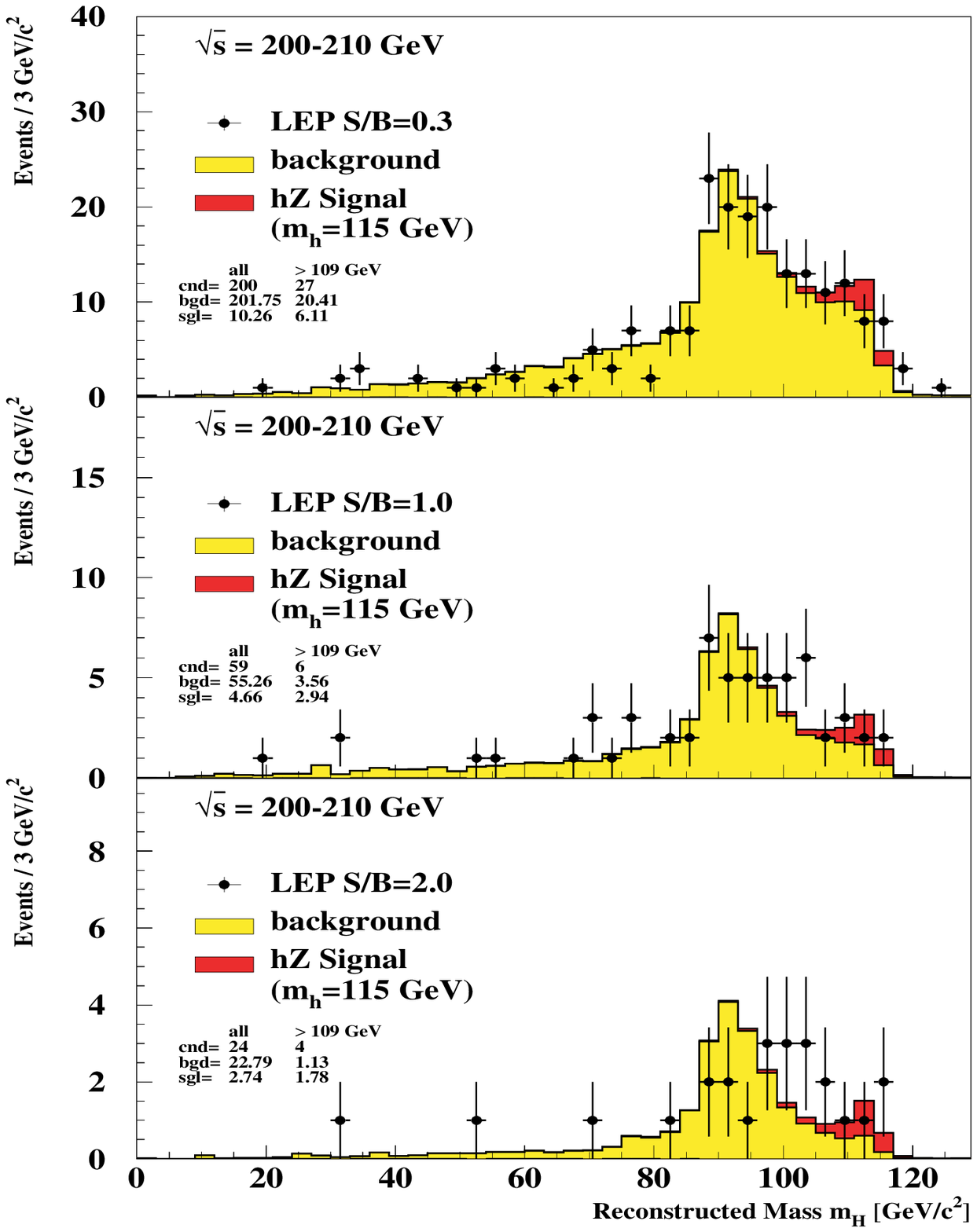,width=1.00\textwidth}
  \caption{Distributions of the reconstructed Higgs mass, $m_{rec} $,
obtained from three special, non-biasing, selections with
increasing signal purity. In the loose/medium/tight selections the
cuts are adjusted in such a way as to obtain, for a Higgs boson of
115 GeV mass, approximately 0.5/1/2 times as many expected signal
as background events in the region $m_{rec} >
109$~GeV\cite{bib-pik}.}
  \label{Fig-masspeak}
\end{figure}

Figure \ref{Fig-masspeak}, taken from the above mentioned
presentation, shows the reconstructed mass distribution of the
background and the signal (on top of the background) with the data
represented by the dots with error bars. It is shown for three
possible selections of the data samples with increasing signal
purity. None of these distributions shows a clear classical
3$\sigma$ excess of data over the expected background and some
physicists claimed that this evidence was not  convincing enough.
However,
 the statistical arguments presented by the LEP Higgs working
 group were not based on these distributions,
  but rather on a sophisticated, though beautiful
 statistical analysis of the data. Two years after the event, when
 the last analysis of the LEP data indicated that the significance
 of a Higgs observation in the vicinity of 115 GeV went down to less than  2$\sigma$ \cite{bib-ADLO-ICHEP},
 it becomes apparent that the LEP Standard Model (SM) Higgs heritage will in fact be a lower
 bound on the mass of the Higgs boson. However, the LEP
 Higgs working group has taught us  powerful and instructive  lessons of
 statistical methods for deriving limits and confidence levels in
 the presence of mass dependent backgrounds from various channels
 and experiments. These lessons will remain with us long after the
 lower bound  becomes outdated.

 In this note we are trying to repeat, in a pedagogical way, this
 LEP Higgs statistical lesson.
 To achieve this we developed a toy Monte Carlo model containing a simulated
 background and signal similar to LEP conditions at $\sqrt s=206$
 GeV.

 Due to mass resolution effects an event with a reconstructed mass
 $m_{rec}$ can originate from any Higgs mass in its vicinity. To
 appreciate the significance of a Higgs candidate with respect to
 some test mass $m_H$ the event weights are introduced in section
 2.

 The weights of all candidate events are summed up  in section 3 to give the
 likelihood of an experiment in order to quantify its signal-like
 nature.

 The probability density function of the likelihood is used in
 section 4 to estimate the probability of an experiment without
 a Higgs boson to fluctuate and give a more signal-like outcome than the observed one
 or the probability of an experiment in the presence of a Higgs signal
 to fluctuate and give a more background-like outcome than the observed one.
  Confidence levels are introduced for that purpose. Later
 on
 in that section, signal confidence levels are defined and used to
 estimate the exclusion sensitivity of an experiment.

 The  current LEP SM Higgs search results are also given,
  allowing the reader to make his own appreciation and
 judgement.

 \section{Event Weights or ``the Spaghetti Higgs Factory" }

\begin{figure}
  \centering
  \epsfig{file=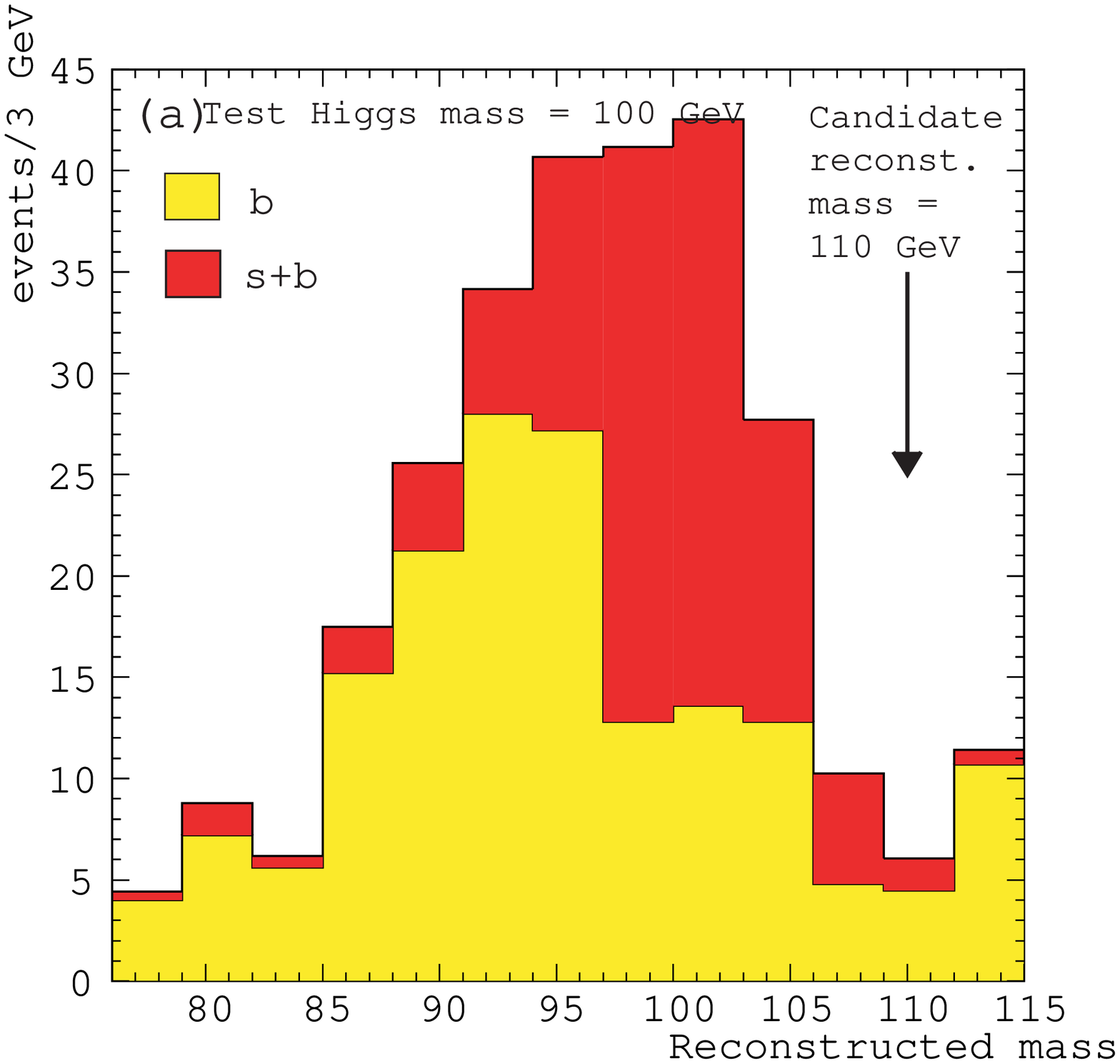,width=0.45\textwidth}
   \epsfig{file=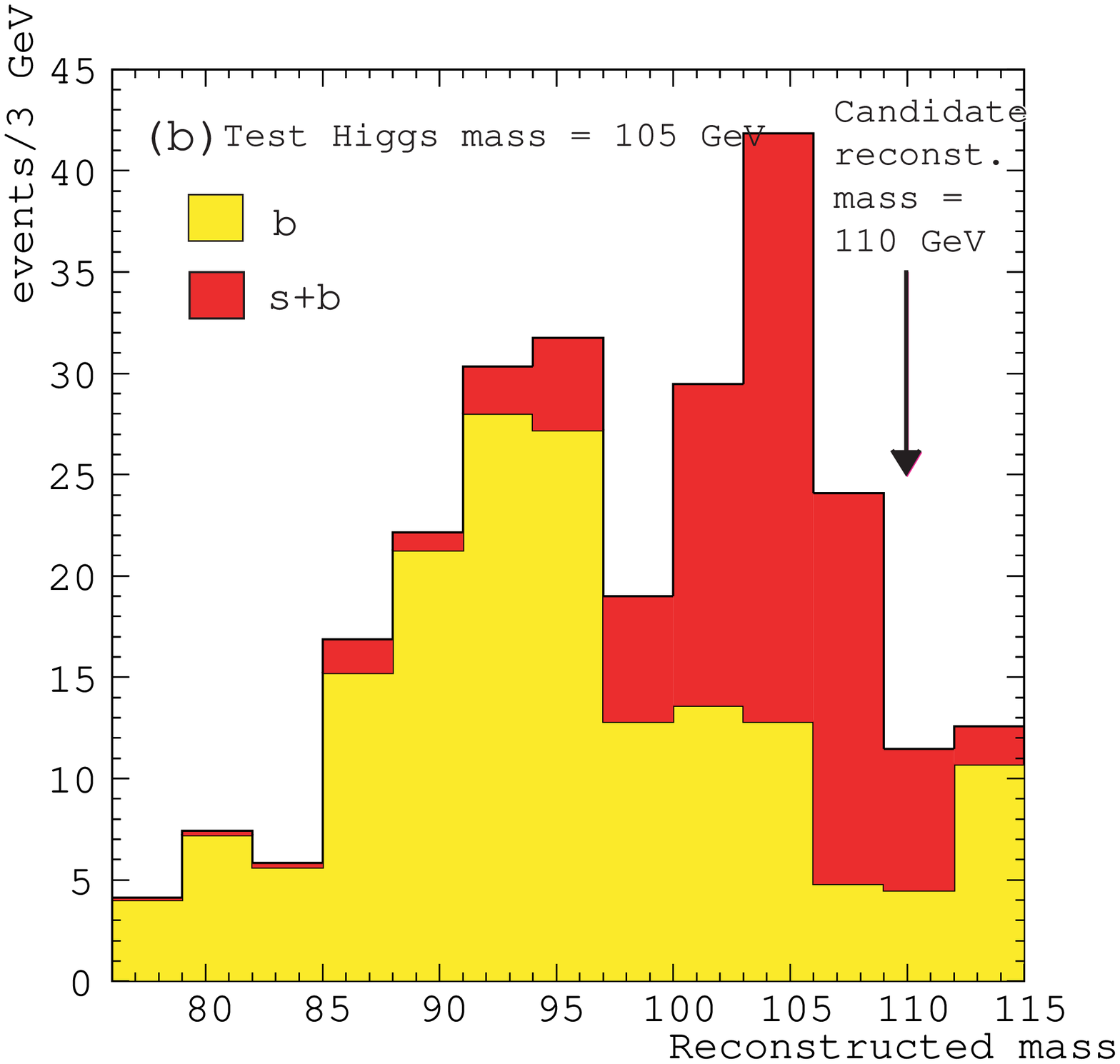,width=0.45\textwidth} \\
  \epsfig{file=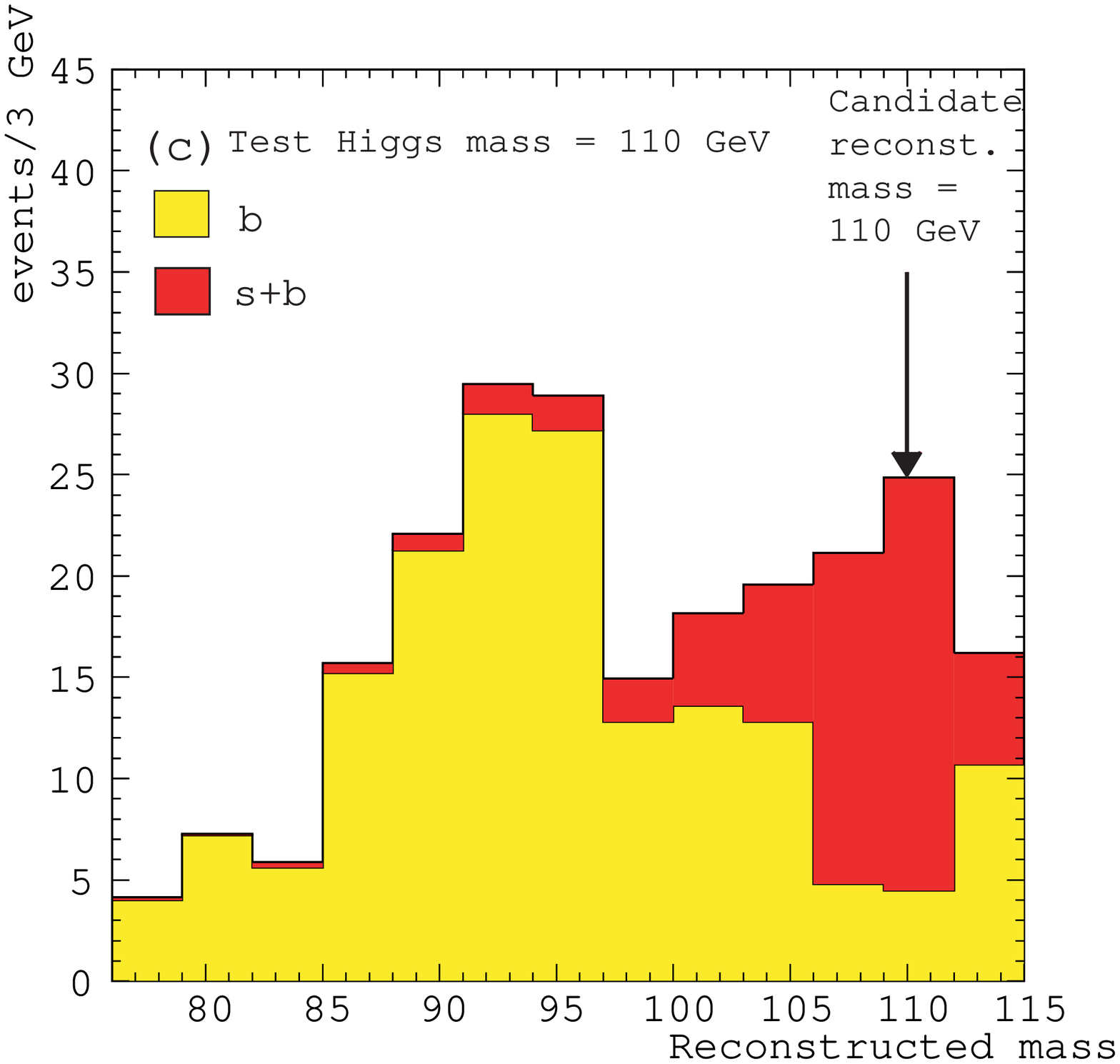,width=0.45\textwidth}
   \epsfig{file=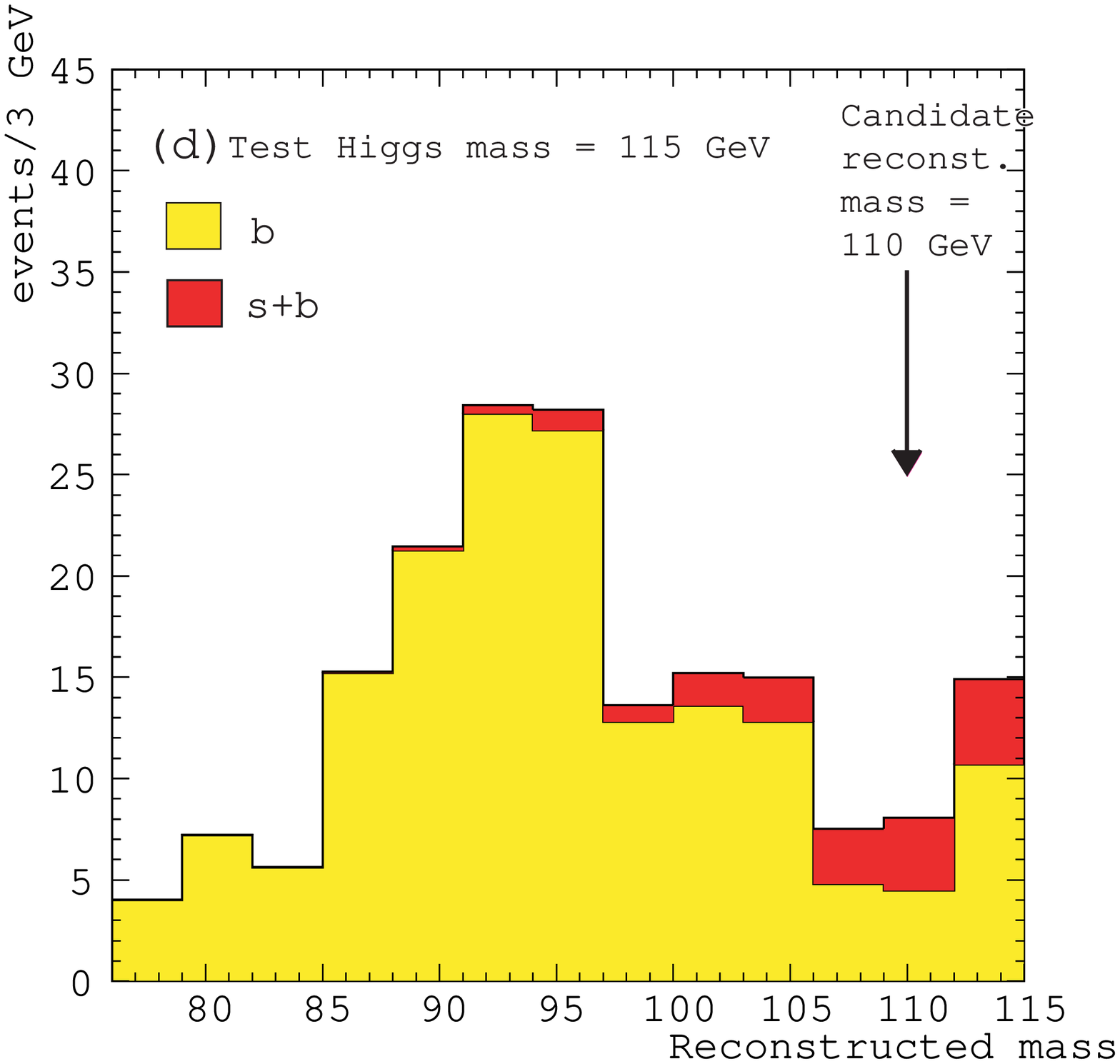,width=0.45\textwidth} \\
     \epsfig{file=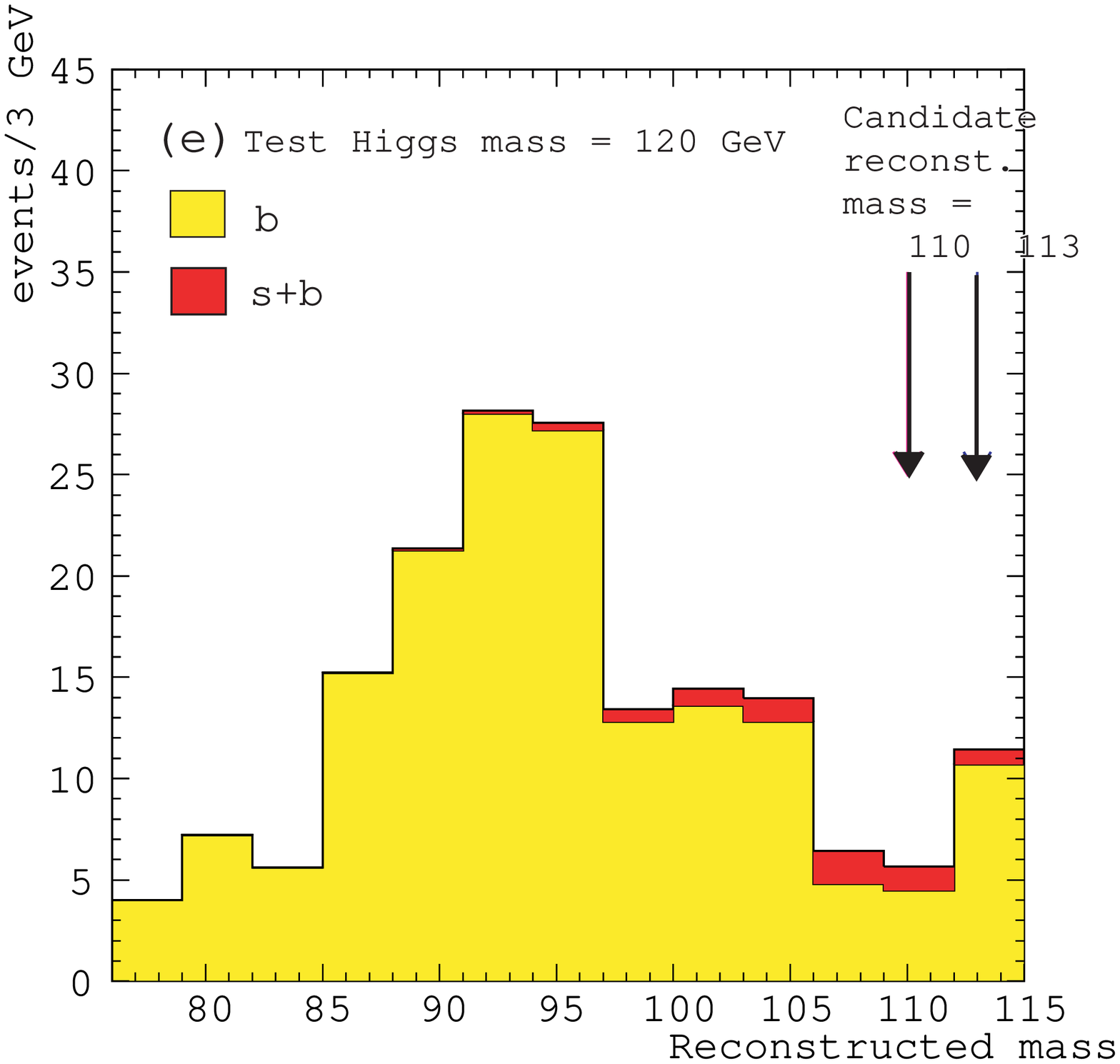,width=0.45\textwidth}
   \epsfig{file=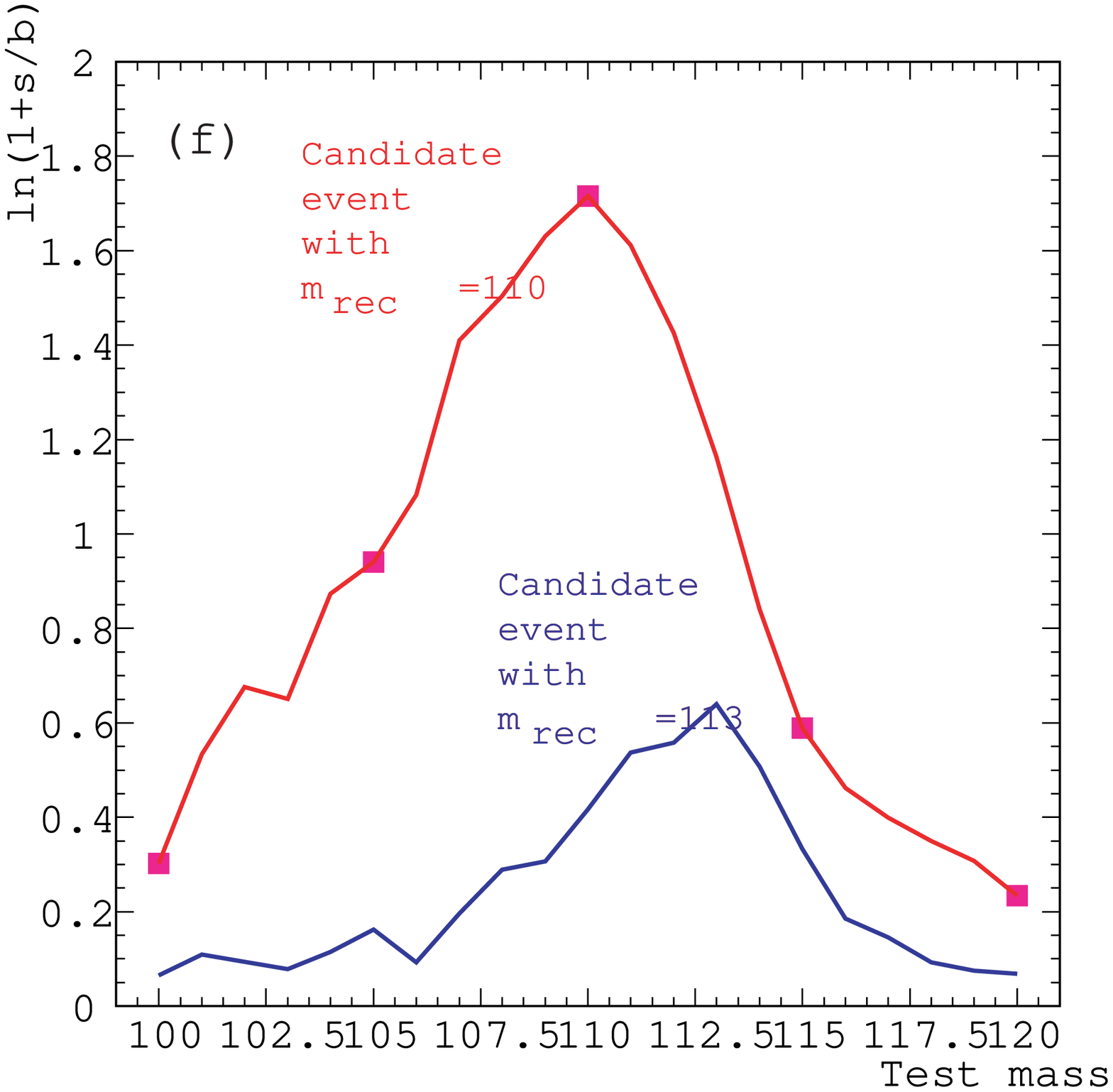,width=0.45\textwidth}
  \caption{\small The "spaghetti" curves represent the evolution of the event weight $\ln(1+s/b)$ with test mass $m_H$.
  The first five plots show the histograms of the  signal (red) on top of the background
  (yellow) for test masses from 100-120 GeV. The arrow represent the location of the candidate reconstructed mass
  at $m_H=110$ GeV. The bottom right plot shows the resulting spaghetti plot for
  a candidate with $m_{rec}=110$ and $113$ GeV.}\label{Fig-spaghetti}
\end{figure}

\begin{figure}
  \centering
  \epsfig{file=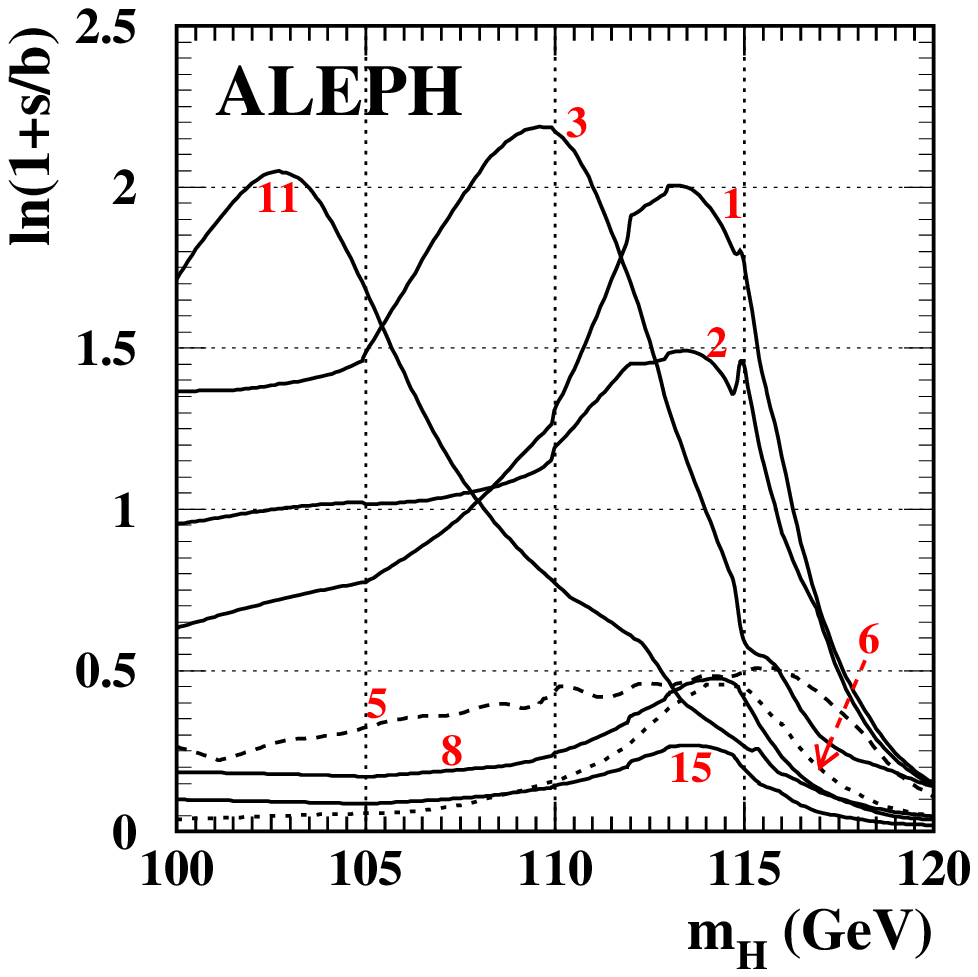,width=0.45\textwidth}
   \epsfig{file=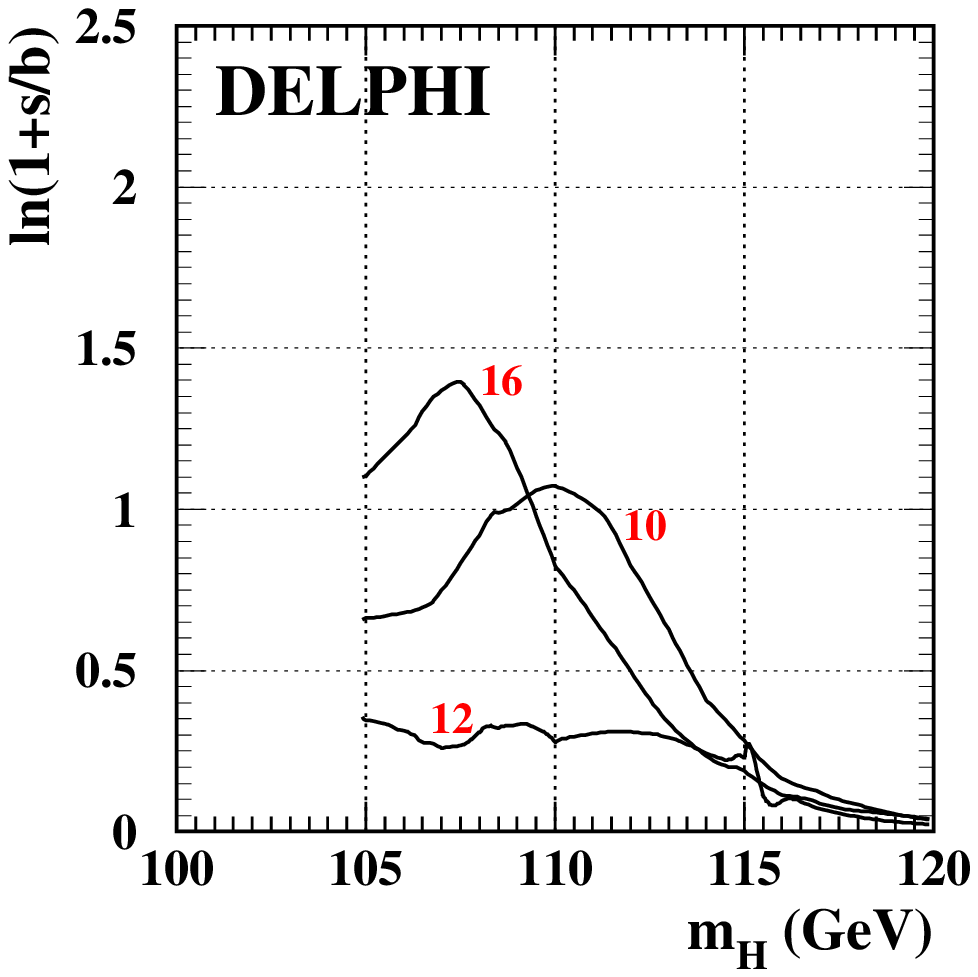,width=0.45\textwidth} \\
  \epsfig{file=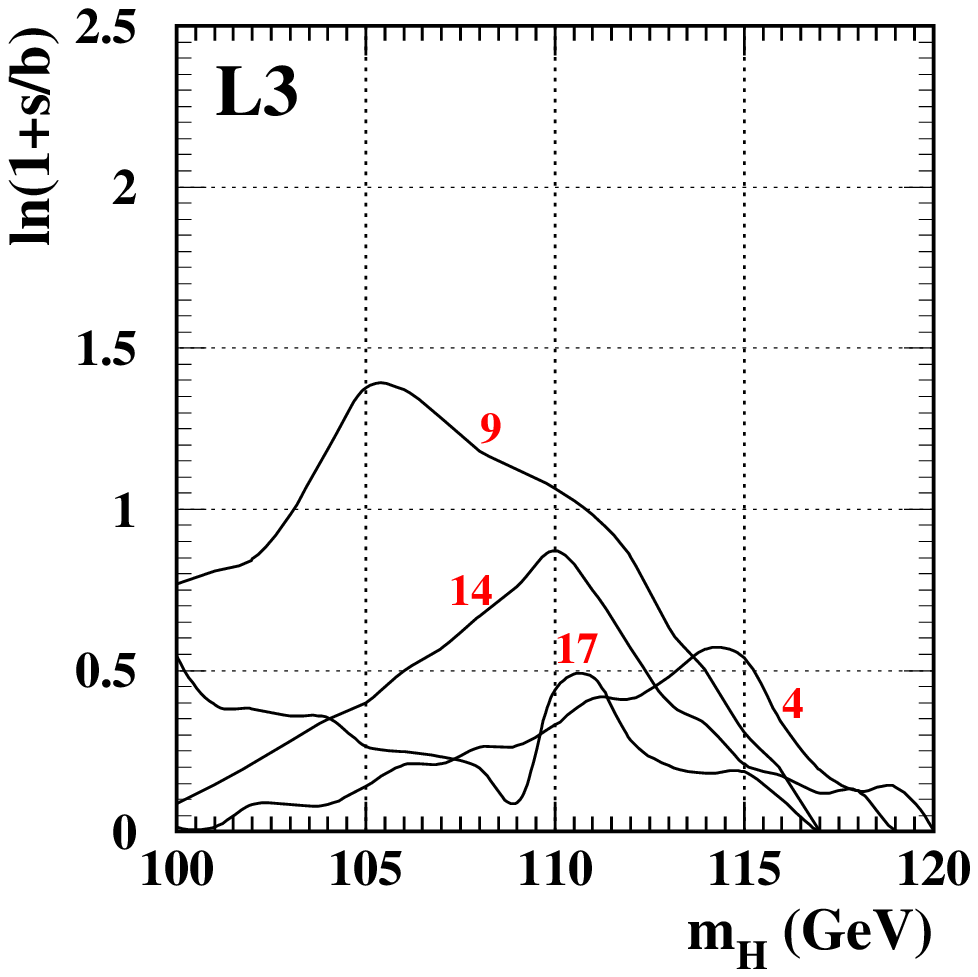,width=0.45\textwidth}
   \epsfig{file=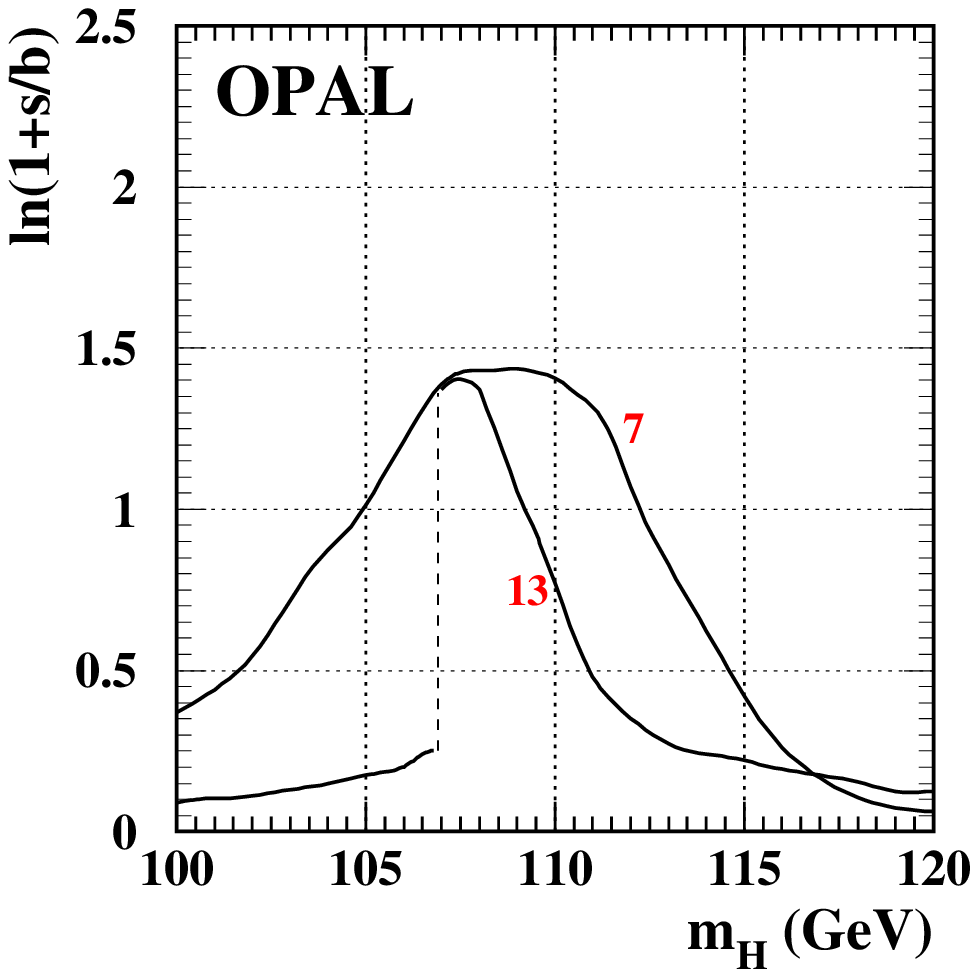,width=0.45\textwidth}

  \caption{{\small
Evolution of the event weight $\ln (1+s/b)$ with test-mass $m_H$
for the events with the largest weight at $m_H =115$~GeV. The
labels correspond to the candidate numbers in the first column of
Table~\ref{tab:event-list}.} The sudden increase in the weight of
the OPAL missing-energy candidate labeled ``13" at $m_H = 107$~GeV
is due to the switching from the low-mass to high-mass
optimization of the search at that mass. A similar increase is
observed in the case of the  L3 four-jet candidate labeled ``17"
which is due to a test-mass dependent attribution of the jet-pairs
to the Z and Higgs bosons.The Figure is taken from
\cite{bib-ADLO-ICHEP}. }\label{Fig-spaghetti-LEP}
\end{figure}

 When does a Higgs candidate event become significant? What can be considered as a leading
 or gold plated Higgs candidate? Due to detector resolution and missing energy carried by
 undetected neutrinos which accompany the Higgs decay products, the reconstructed
 mass of a hypothetical Higgs candidate $m_{rec}$,
 is not necessarily its physical mass, $m_H$. One therefore defines a weight
 to quantify the significance of a Higgs candidate with $m_{rec}$
 with respect to a hypothetical Higgs with $m_H$. This weight is
 given by
  $\ln(1+\frac{s(m_H,m_{rec})}{b(m_{rec})})$ (see next section)
 where $s(m_H,m_{rec})$ is the expected number of signal events
 from an hypothetical Higgs boson with a test mass $m_H$ in the
 vicinity of $m_{rec}$.\footnote{The histograms and consequently the weights are
  binned taking into account the experimental resolution and the statistics.}
   This procedure is
 illustrated in a series of plots (Figure \ref{Fig-spaghetti}).
 In this example we assume there is an event candidate with a
 reconstructed mass, $m_{rec}=110$ GeV. We then calculate the
 event weight for  hypothetical Higgs test masses in the range   100--120 GeV
 (Figs \ref{Fig-spaghetti}a-e). The resulting weights are shown in
 the last figure of this series (Fig. \ref{Fig-spaghetti}e). When
 connecting all the weights with a continuous line, one gets what
 has become to be known as the ``spaghetti plot". Also shown is the
 spaghetti plot for a candidate event with $m_{rec}=113$ GeV.
 As expected, the highest weight is achieved when the reconstructed
 mass and the hypothetical test Higgs mass coincide.
 The 17 candidates with the highest weight at a test mass,
 $m_H=115$ GeV are listed in Table \ref{tab:event-list} and their
 corresponding spaghetti plots are shown in Figure
 \ref{Fig-spaghetti-LEP} (taken from \cite{bib-ADLO-ICHEP}).

 \begin{table}[htb]
\begin{center}
\begin{tabular}{|c|lcc|cc|}
\hline
   & Expt&  $E_{cm}$ &  Decay channel &   $m_{rec}$ (GeV) &  $\ln (1+s/b)$ \\
   &     &           &                &                     &  at 115 GeV    \\
\hline\hline
1&  ALEPH  &  206.6  &  4-jet         &   114.1             & 1.76           \\
2&  ALEPH  &  206.6  &  4-jet         &   114.4             & 1.44           \\
3&  ALEPH  &  206.4  &  4-jet         &   109.9             & 0.59           \\
4&  L3     &  206.4  &  E-miss        &   115.0             & 0.53           \\
5&  ALEPH  &  205.1  &  Lept          &   117.3             & 0.49           \\
6&  ALEPH  &  206.5  &  Taus          &   115.2             & 0.45           \\
7&  OPAL   &  206.4  &  4-jet         &   111.2             & 0.43           \\
8&  ALEPH  &  206.4  &  4-jet         &   114.4             & 0.41           \\
9&  L3     &  206.4  &  4-jet         &   108.3             & 0.30           \\
10& DELPHI &  206.6  &  4-jet         &   110.7             & 0.28           \\
11& ALEPH  &  207.4  &  4-jet         &   102.8             & 0.27           \\
12& DELPHI &  206.6  &  4-jet         &    97.4             & 0.23           \\
13& OPAL   &  201.5  &  E-miss        &   108.2              & 0.22           \\
14& L3     &  206.4  &  E-miss        &   110.1             & 0.21           \\
15& ALEPH  &  206.5  &  4-jet         &   114.2             & 0.19           \\
16& DELPHI &  206.6  &  4-jet         &   108.2             & 0.19           \\
17& L3     &  206.6  &  4-jet         &   109.6             & 0.18           \\
\hline
\end{tabular}
\caption{\small Properties of the candidates  with the highest
weight at $m_H =115$~GeV. Table is taken from
\cite{bib-ADLO-ICHEP}. \label{tab:event-list}}
\end{center}
\end{table}

\section{Understanding Likelihood  Plots}

At the end of the day, LEP is one big experiment with one
experimental result. An experimental result is in this sense  a
configuration of data events that agree to some level the
expectation from either a pure \textbf{background (\textit{b})}
hypothesis or \textbf{signal plus background (\textit{s+b})}
hypothesis (at some Higgs mass). Here we illustrate how the
likelihood ratio is used to rank an experimental result between
either being \textit{b}-like or \textit{s}+\textit{b}-like.

The first step in telling a \textit{b}-like from an
\textit{s}+\textit{b}-like result is to construct a discriminator.
Such a discriminator could be the reconstructed mass (obviously a
peak at the reconstructed mass on top of the background will
indicate a ``signal observation"). It could also be a 2-D
discriminator, where the other discriminating variable is, for
example, the b-tag content of an event. The Higgs, being the
generator of particle masses, decays dominantly to b-quarks in
this mass range. In our toy model, we use the reconstructed mass
as a discriminator.

Once the discriminator is defined, we divide it into bins,
$i=1,2,.....,N_{bins}$  each containing $N_i$ observed candidates.
The likelihood ratio $-2\ln Q(m_H)$ tells us how much the outcome
of an experiment is signal-like \cite{bib-AlexRead}. It is given
by
\begin{equation}
Q=\frac{P_{Poisson}(Data|s+b)}{P_{Poisson}(Data|b)}=\frac{L(s+b)}{L(b)}
=\frac{\exp(-(s_{TOT}+b_{TOT}))}{\exp(-b_{TOT})}\prod_{i=1}^{N_{bins}}\left({\frac{s_i+b_i}{b_i}}\right)^{N_i},
\end{equation}
which is easily simplified to
\begin{equation}
-2\ln
Q(m_H)=2s_{TOT}-2\sum_{i=1}^{N_{bins}}N_i\ln\left(1+\frac{s_i(m_H)}{b_i}\right),
\label{eq-Q}
\end{equation}
which is a weighted sum of all the observed events.

Next we demonstrate how to construct the Probability Density
Function (\textbf{p.d.f.}) of the likelihood for an ensemble of
background and signal+background experiments.

\begin{figure}
  \centering
  \epsfig{file=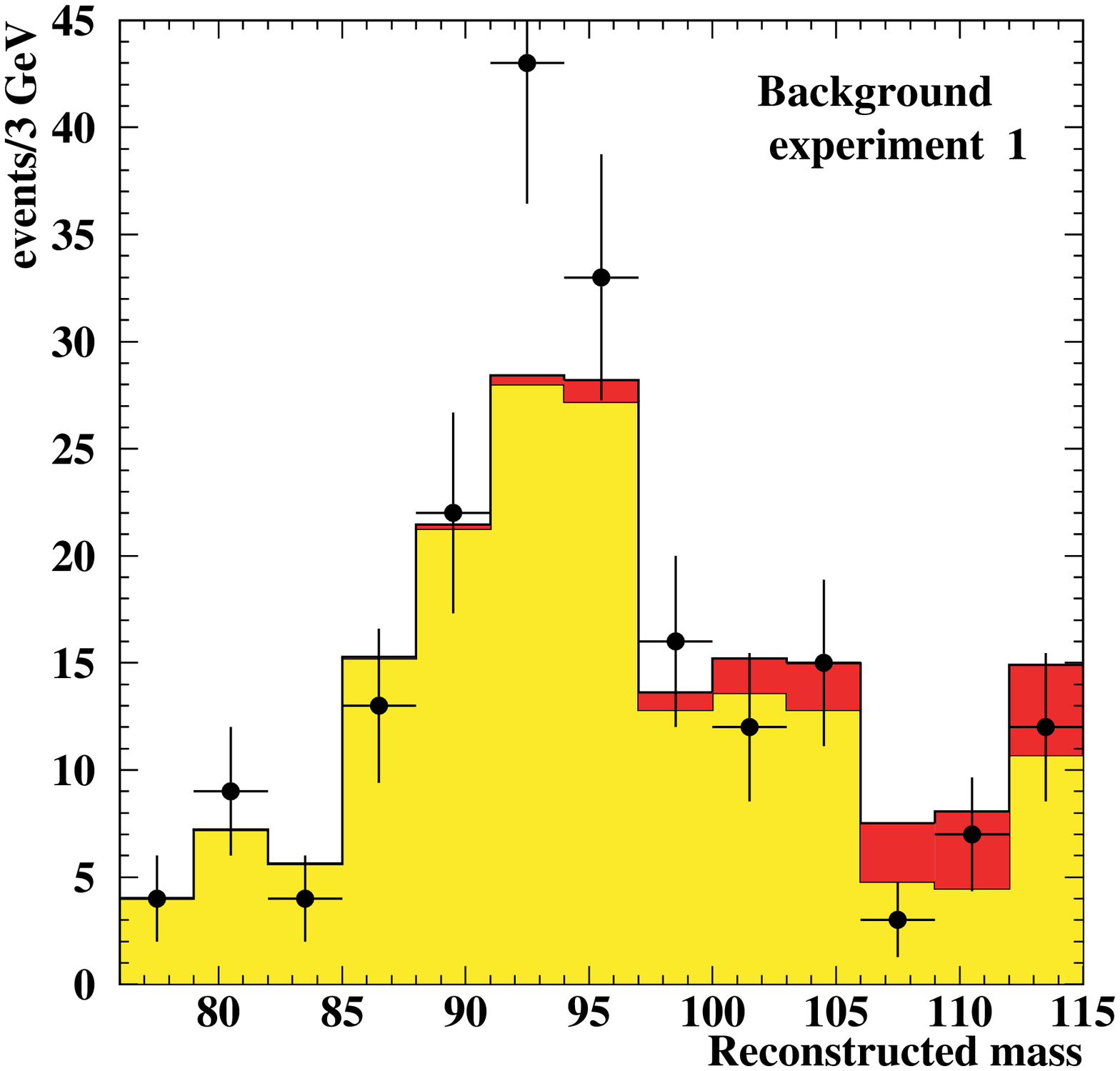,width=0.43\textwidth}
   \epsfig{file=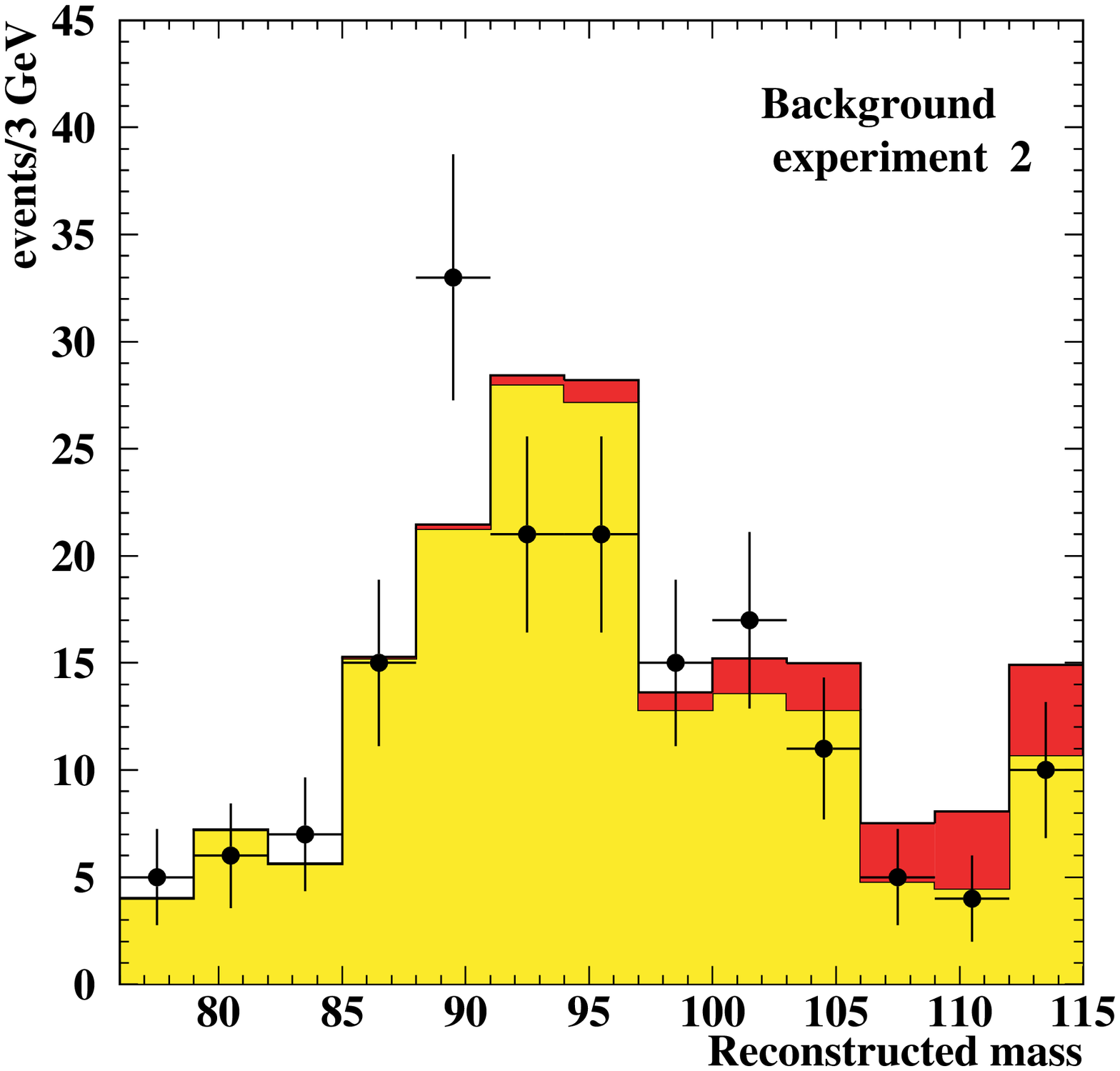,width=0.43\textwidth} \\
  \epsfig{file=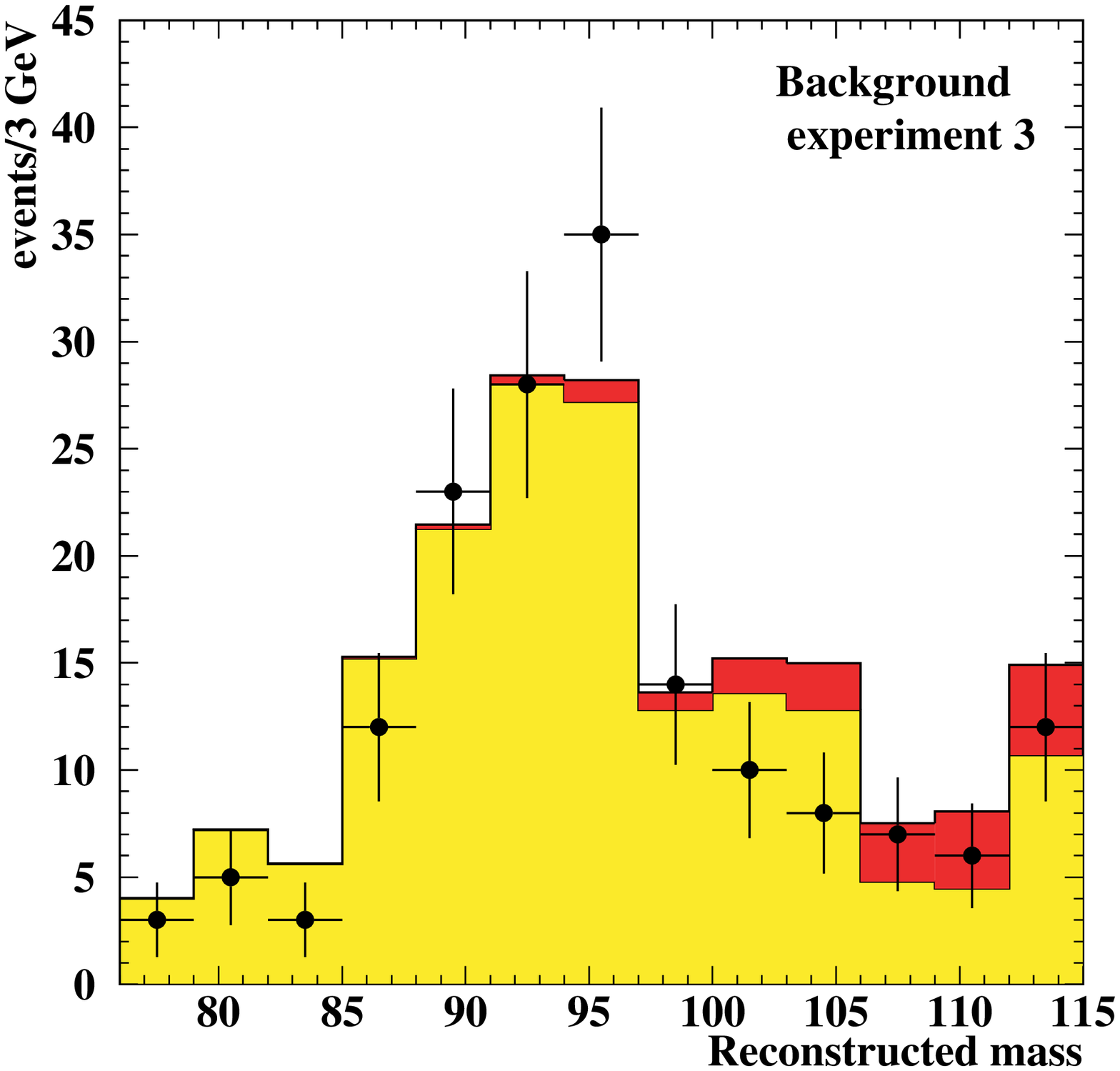,width=0.43\textwidth}
   \epsfig{file=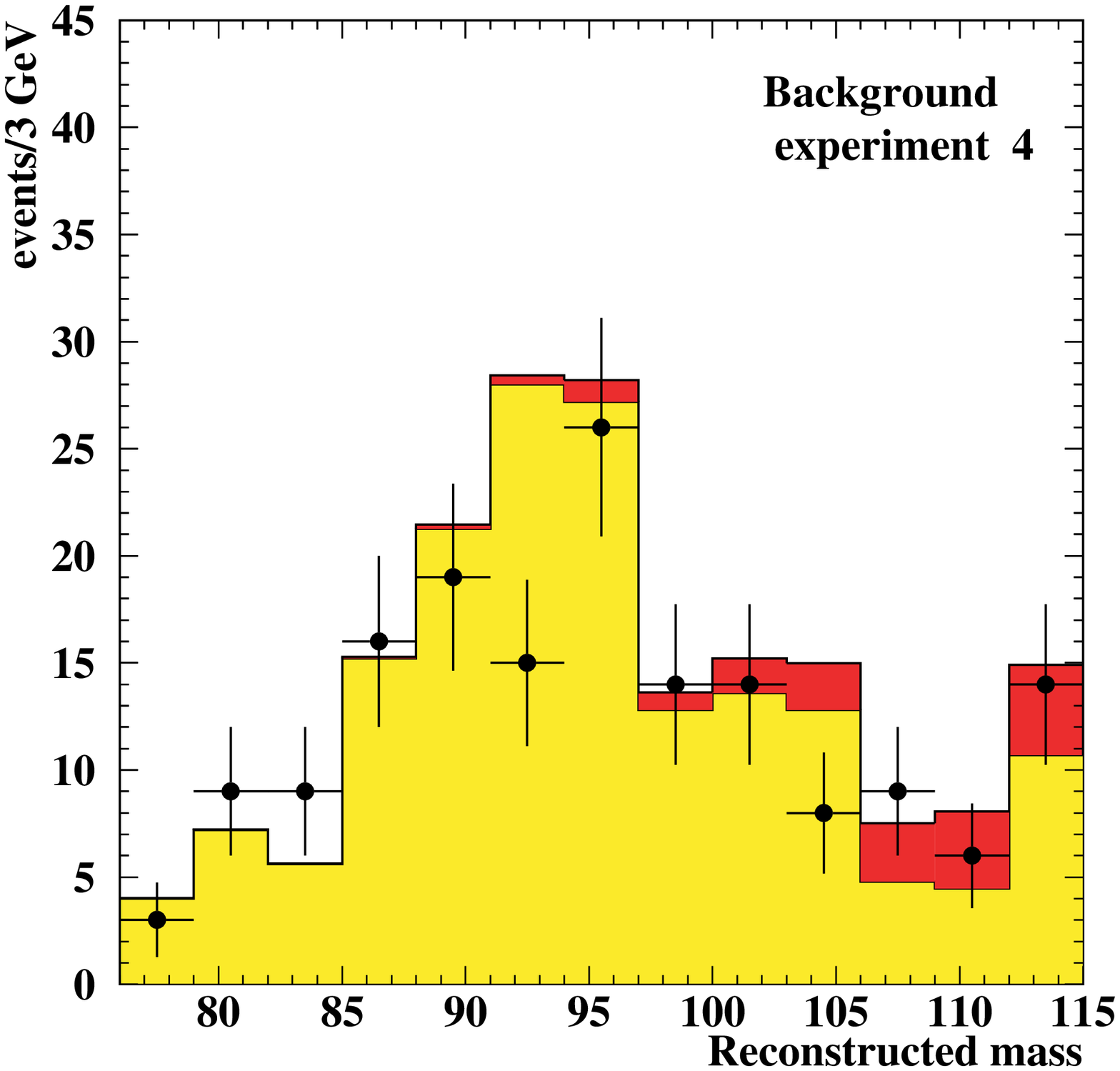,width=0.43\textwidth} \\
     \epsfig{file=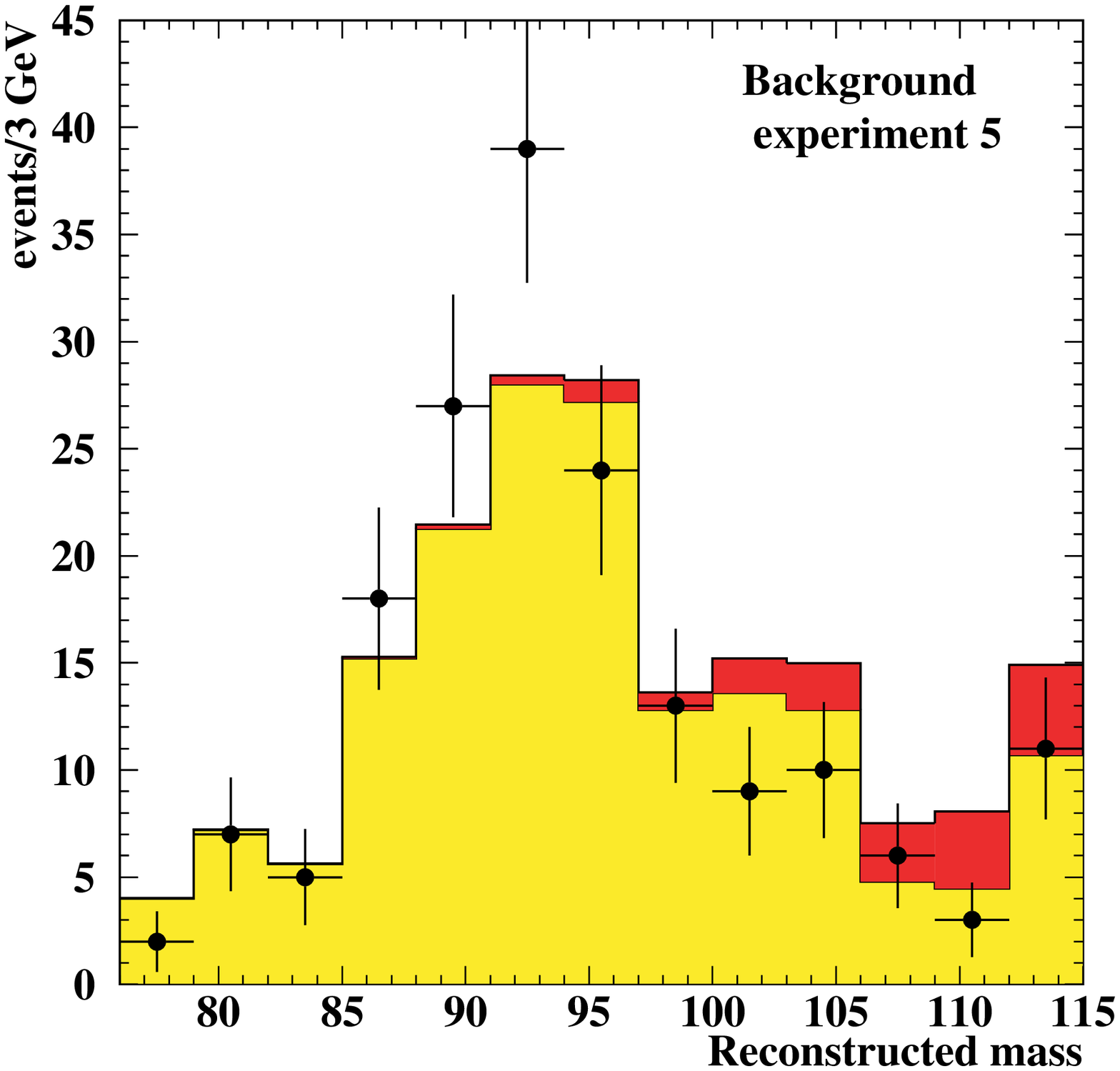,width=0.43\textwidth}
       \epsfig{file=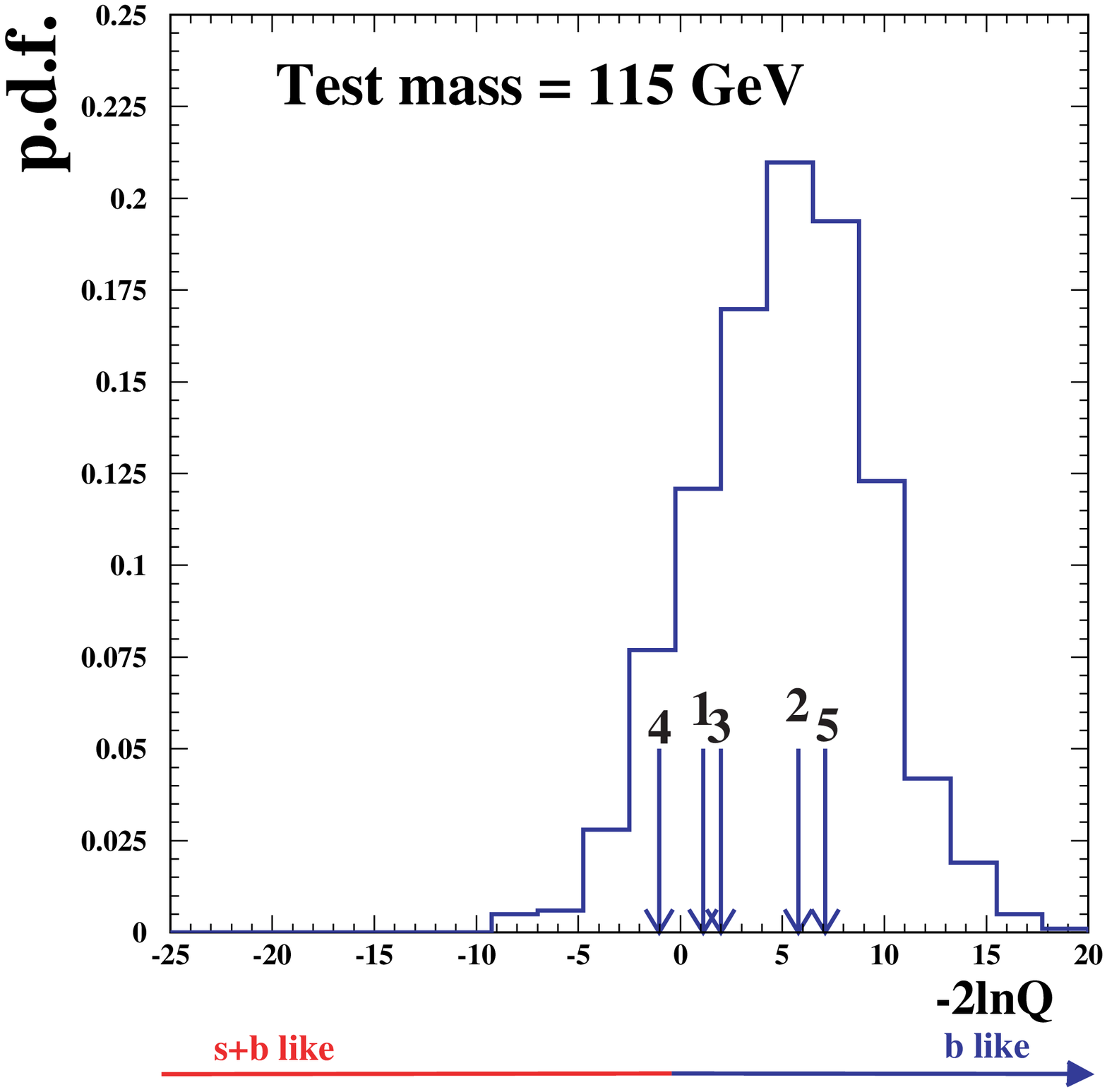,width=0.43\textwidth}

  \caption{\small The first five plots show the histograms of the 115 GeV Higgs signal (red) on top of the background
  (yellow) for a Higgs test mass of 115 GeV. The dots with error bars represent
  a simulated data of random background only experiments.
  Each experiment yields a likelihood represented by the arrows in
  the bottom right plot. By performing an infinite number of
  background only experiments one gets the probability density function
  shown in the bottom right plot.}\label{Fig-bgexperiments}
\end{figure}

The histograms in Figure \ref{Fig-bgexperiments} show the expected
background and 115 GeV Higgs signal (on top of the background)
distributions of the discriminating variable, here taken to be the
reconstructed mass. We then show five possible outcomes of
experiments, all generated without the signal. The experiments are
numbered 1--5. For each experiment we use the observed events
configuration ($N_i$) and the hypothetical Higgs test mass $m_H$
to calculate the likelihood $-2\ln Q(m_H)$ following equation
\ref{eq-Q}. The resulting likelihoods are shown at the bottom
right plot. Only experiment 4 has some excess of events at the
high mass region which results in an \textit{s}+\textit{b}-like
likelihood. The other 4 experiments  result in a \textit{b}-like
likelihood. The likelihood p.d.f. is the histogram generated when
performing a large number of background only experiments. As one
can see (from simple areas considerations), in this ``typical" toy
example , the probability for a \textit{b}-only experiments to
give a \textit{s}+\textit{b}-like likelihood is about 15\% (not so
small...)\footnote{Later, in this note, we will see that this
probability is $1-CL_b$, where $CL_b$ is the background
confidence.}.

\begin{figure}
  \centering
  \epsfig{file=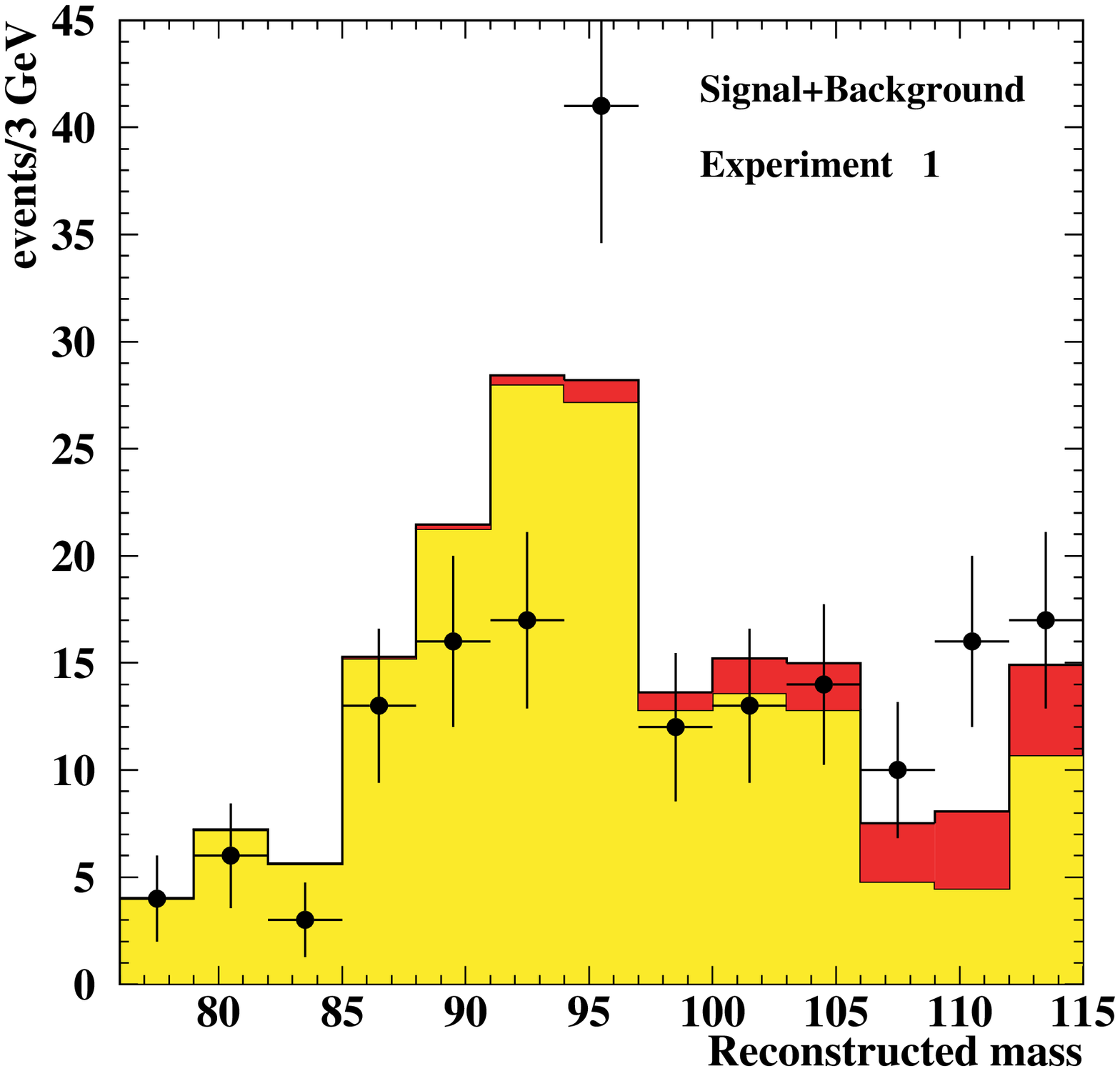,width=0.43\textwidth}
   \epsfig{file=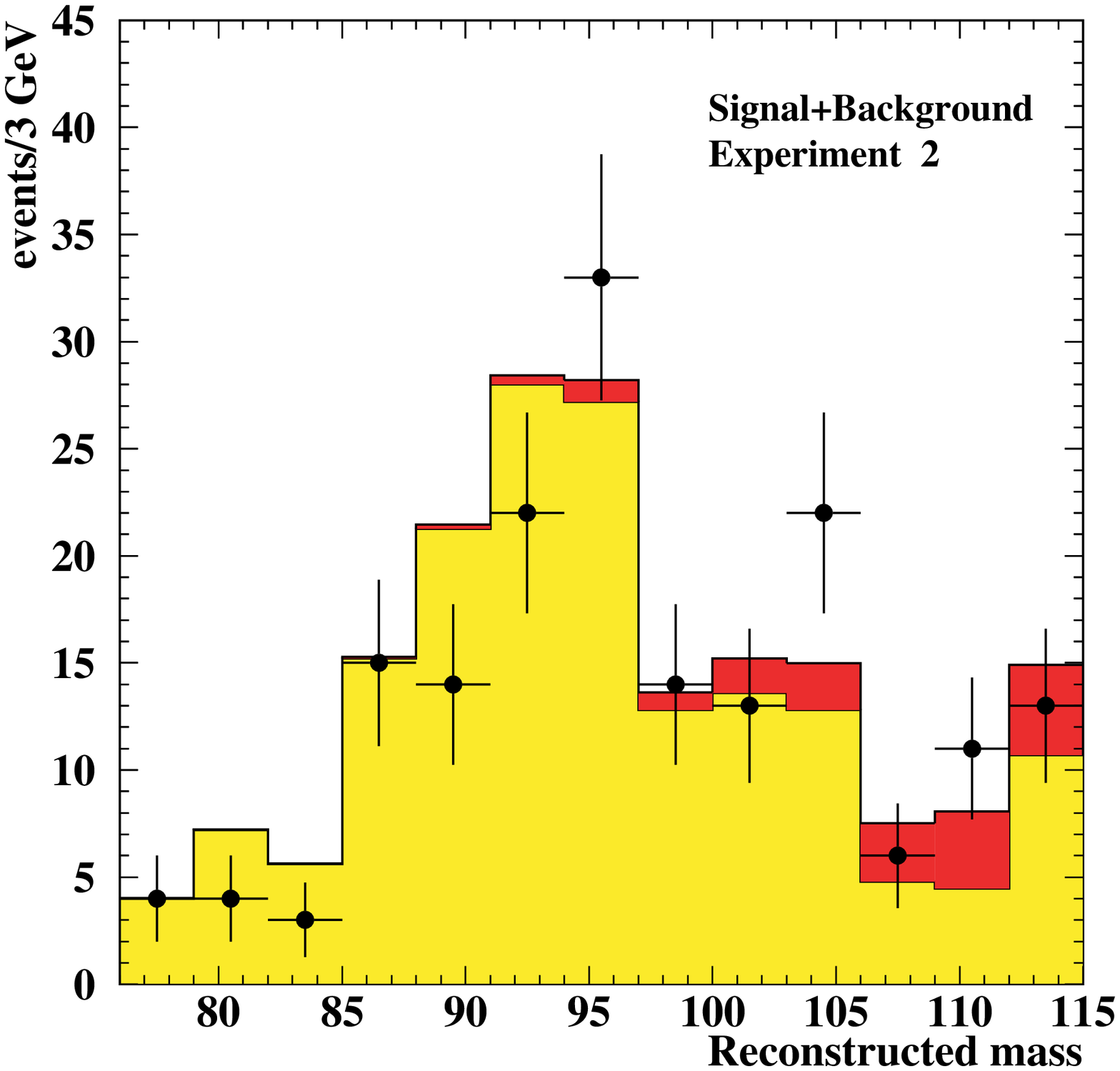,width=0.43\textwidth} \\
  \epsfig{file=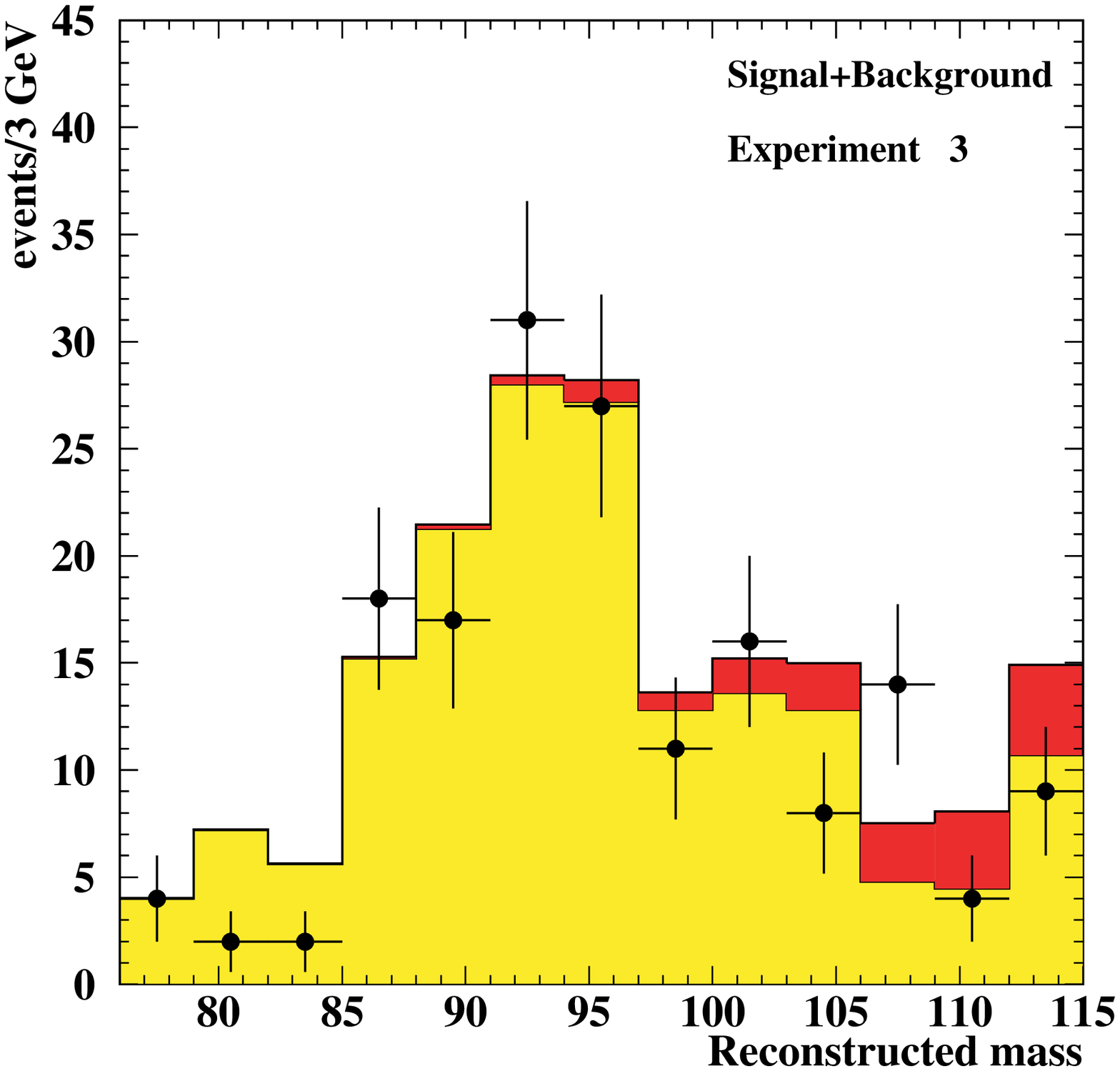,width=0.43\textwidth}
   \epsfig{file=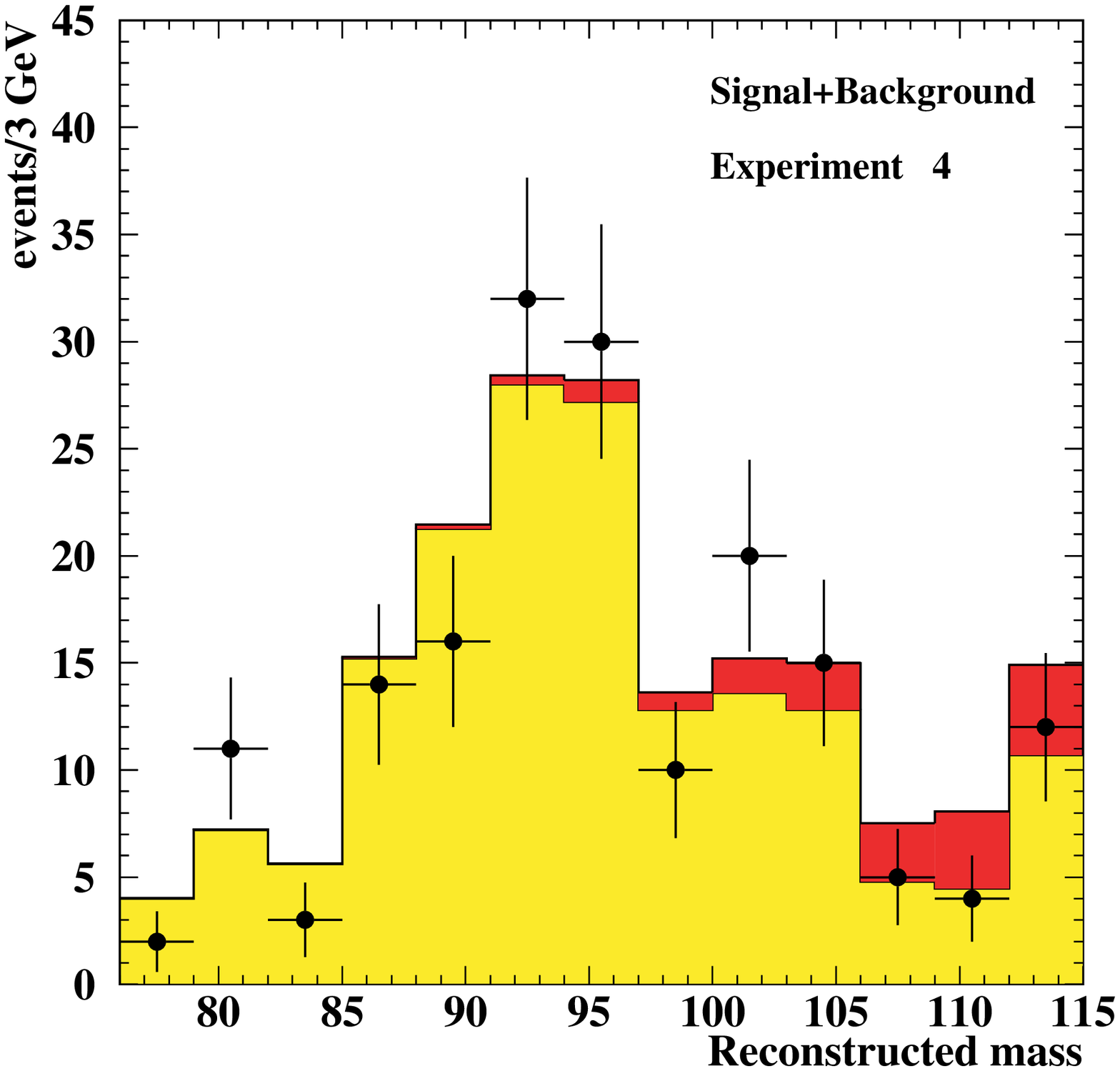,width=0.43\textwidth} \\
     \epsfig{file=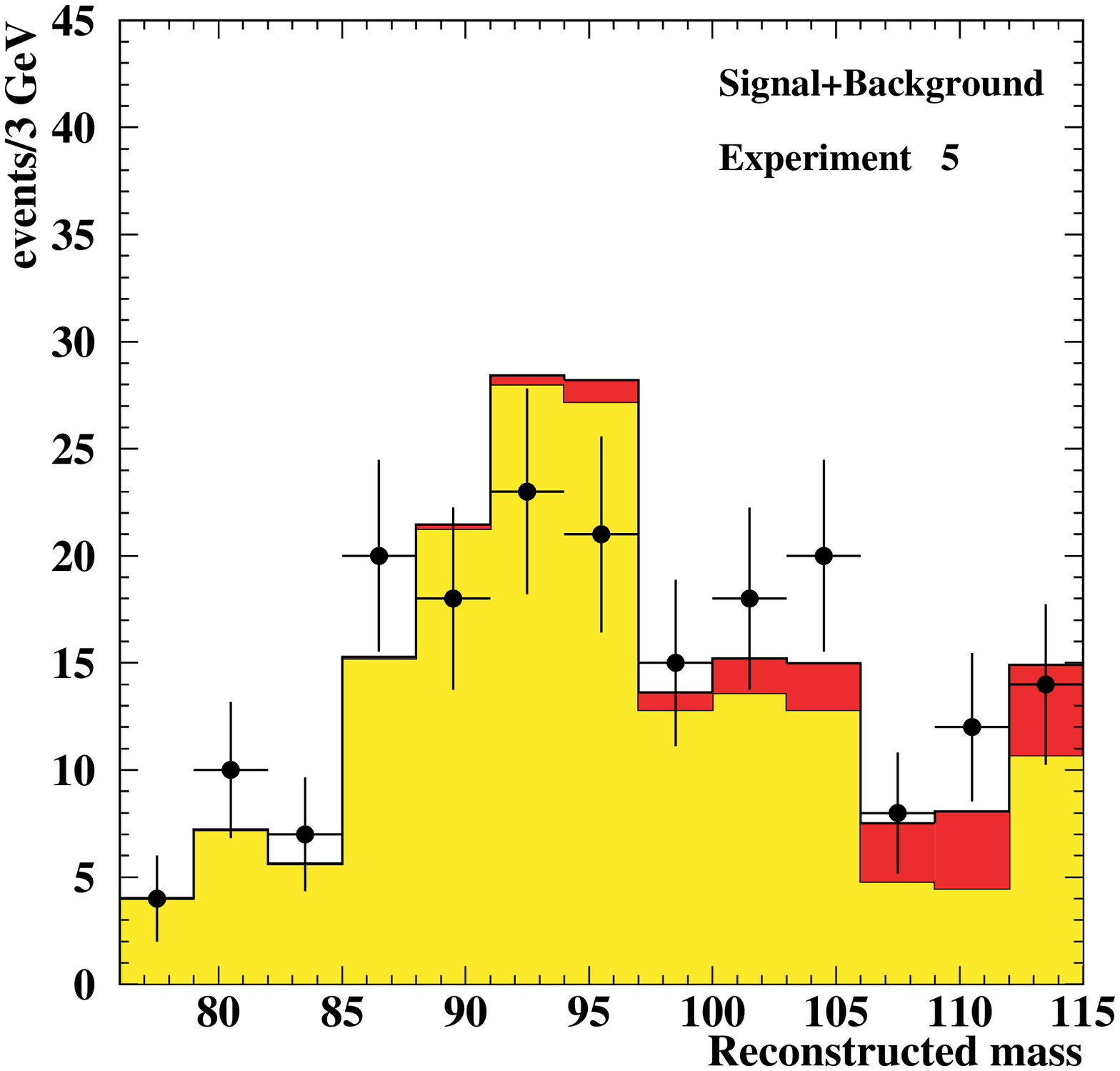,width=0.43\textwidth}
       \epsfig{file=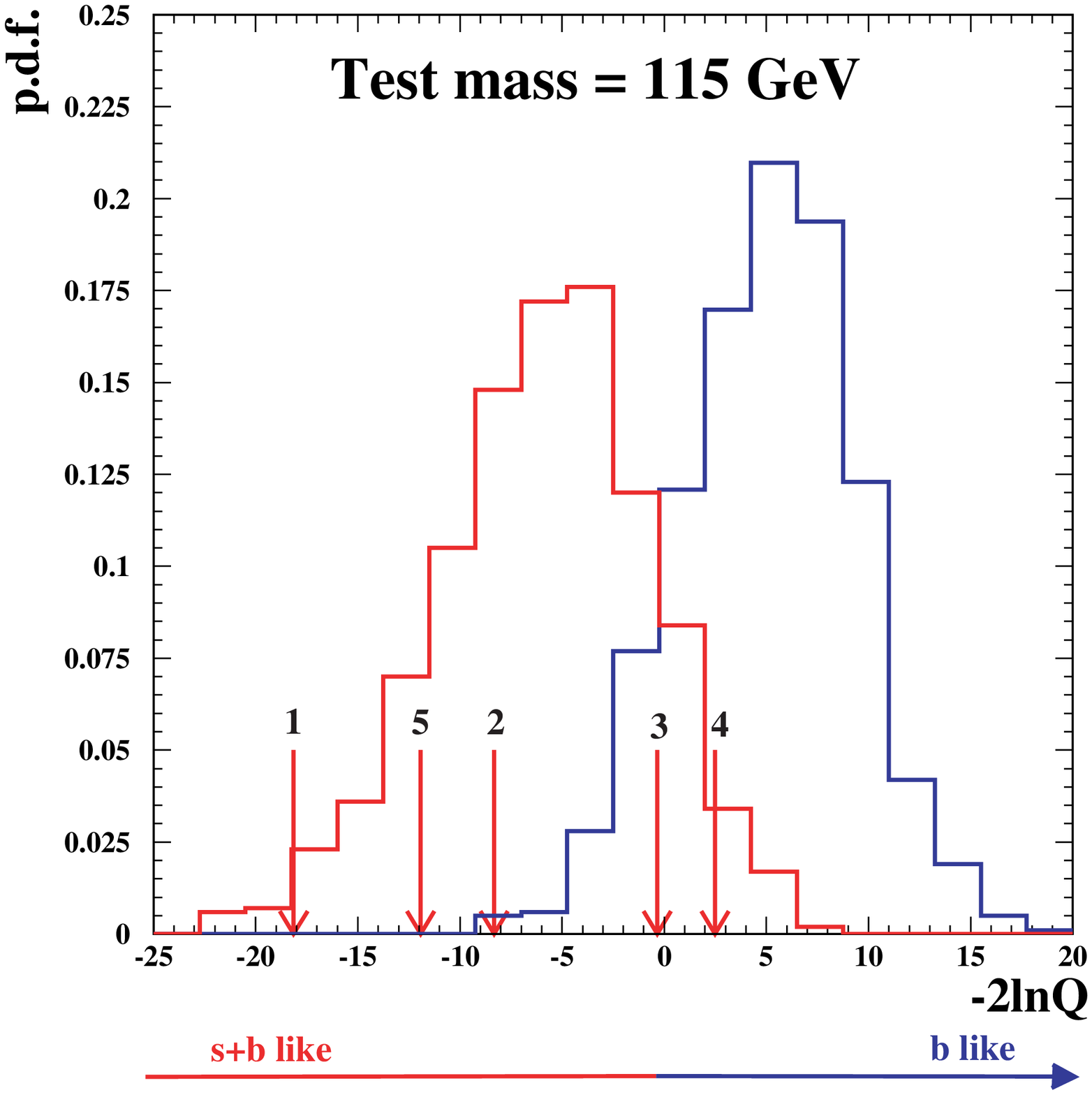,width=0.43\textwidth}

  \caption{\small The first five plots show the histograms of the  115 GeV Higgs signal (red) on top of the background
  (yellow) for a Higgs test mass of 115 GeV. The dots with error bars represent
   simulated data of random signal+background  experiments.
  Each experiment yields a likelihood represented by the arrows in
  the bottom right plot. By performing a large number of
  signal+background only experiments one gets the probability
  density
  function(p.d.f.)
  shown in the bottom right plot. Also shown is the background only p.d.f.}\label{Fig-sexperiments}
\end{figure}

The same procedure was repeated for signal+background experiments
(see Figure \ref{Fig-sexperiments}). The bottom right plot  shows
the likelihood of the five numbered \textit{s}+\textit{b}
experiments. As one can see, experiments  3 and 4 gave a
background-like likelihood. Repeating the \textit{s}+\textit{b}
experiments a large number of times yielded another p.d.f. from
which we see that in this specific toy example, there is about a
20\% probability for a 115 GeV signal to give a \textit{b}-like
event configuration\footnote{Later in this note, we will see that
this probability is the signal+background confidence,
$CL_{s+b}$.}. The p.d.f's of the \textit{s}+\textit{b} and
\textit{b}-only experiments give us an indication of the
discriminating power of the likelihood.

\begin{figure}
  \centering
  \epsfig{file=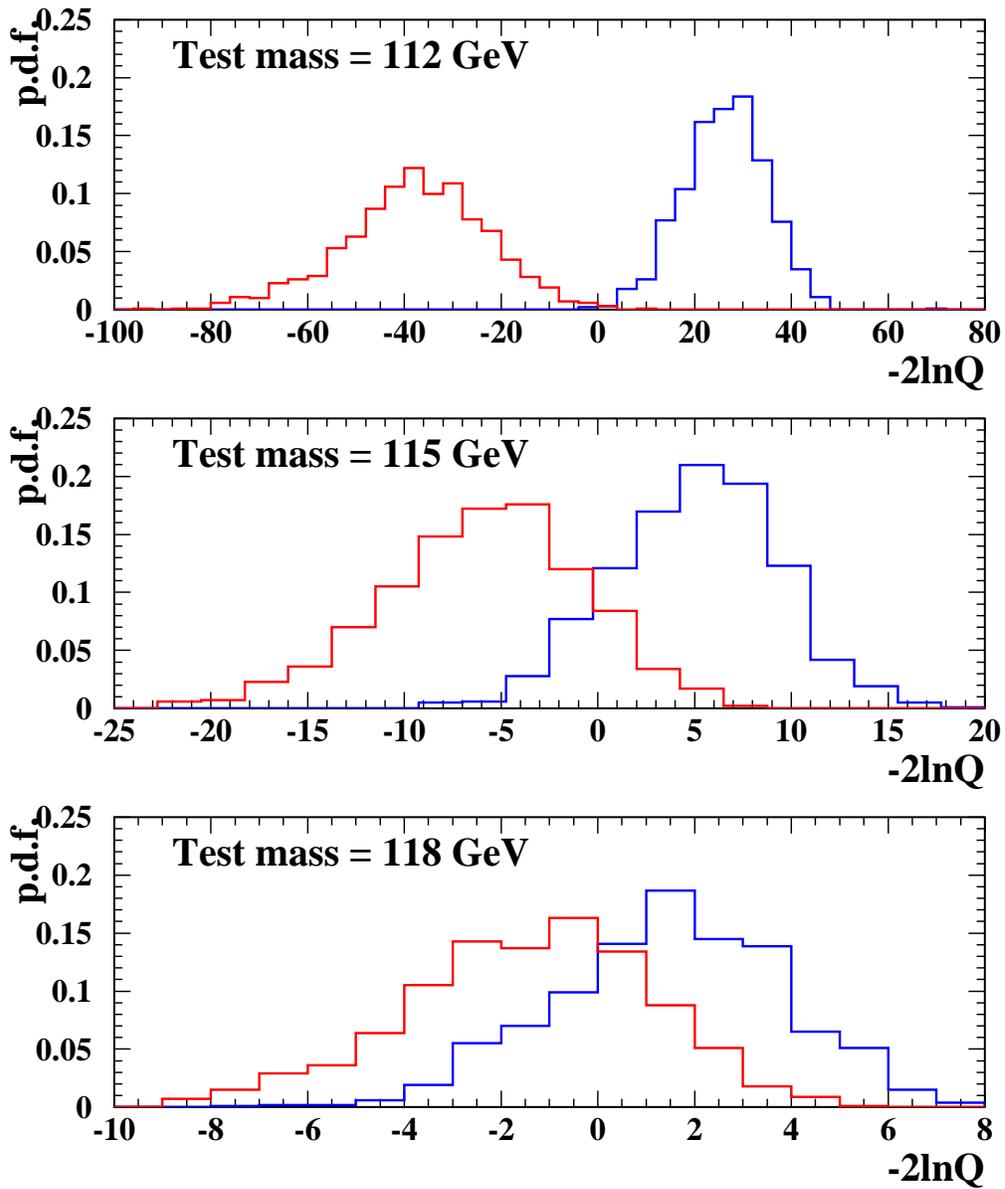,width=1.00\textwidth}

  \caption{The separation between the Signal and the Background for various Higgs masses is
  shown by their likelihood p.d.f's. }\label{Fig-discriminators}
\end{figure}

\begin{figure}
  \centering
  \epsfig{file=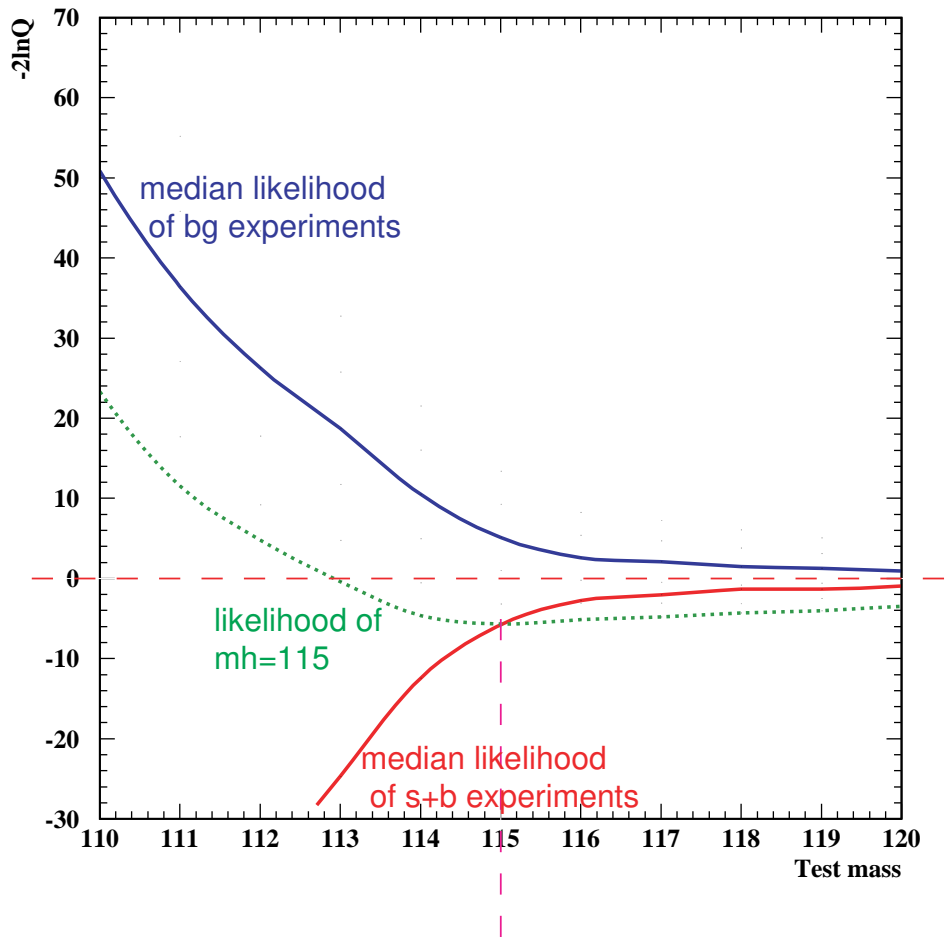,width=1.00\textwidth}

  \caption{Median expected likelihood of background-only (blue), signal(at the test mass $m_H$)+background (red) and a
  fake 115 GeV Higgs signal (green) as a function of the test Higgs mass $m_H$.}\label{Fig-median}
\end{figure}

This is illustrated in Figure \ref{Fig-discriminators} where the
p.d.f's of \textit{b}-only and \textit{s}+\textit{b} experiments
are shown for $m_H=112,115$ and $118$ GeV. For the lighter Higgs,
the production cross section is high allowing a good separation
between signal and background. For the heavier Higgs the signal
production cross section is very low resulting in a weak
separation. The dependence of the likelihood on the hypothetical
test Higgs mass is illustrated in Figure \ref{Fig-median} where
the median of the p.d.f. distributions is shown for
\textit{b}-only and \textit{s}+\textit{b} experiments. As the
Higgs test mass increases the signal and background separation
power decreases resulting in a reduced discovery potential.
 This median can serve as a reference  measure to the signal-like
 nature of the experimental result. To demonstrate this,
 we planted into our background Monte Carlo sample a 115 GeV Higgs boson signal.
We then generated the p.d.f. of this fake signal likelihood. The
resulting p.d.f. median is also shown in Figure \ref{Fig-median}.
One can see, as expected,  that this likelihood reaches a broad
minimum at a test mass in the vicinity of $m_H=115$ GeV.
Naturally, this likelihood coincides with that of the
\textit{s}+\textit{b} experiments for a test mass of 115 GeV.

\begin{figure}
  \centering
  \epsfig{file=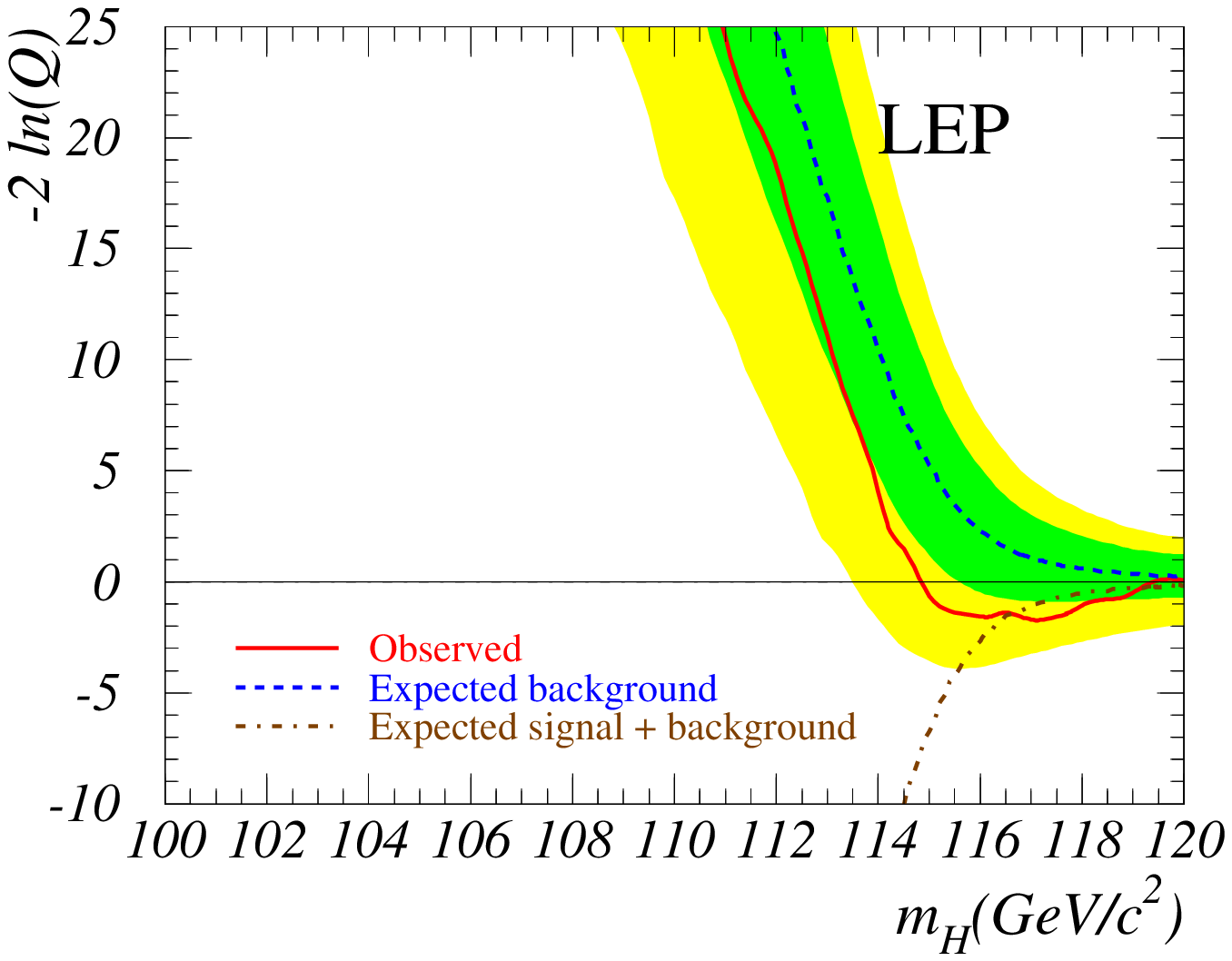,width=1.00\textwidth}
 \caption{ Observed and expected behavior of the likelihood $-2\ln Q$ as a function of the test-mass
$m_H $ for combined LEP experiments. The solid/red line represents
the observation; the dashed/dash-dotted lines show the median
background/signal+background expectations. The dark/green and
light/yellow shaded bands represent the 1 and 2 $\sigma$
probability bands about the median background expectation
\cite{bib-ADLO-ICHEP}.}\label{Fig-likelihood-LEP}
\end{figure}

\begin{figure}
  \centering
  \epsfig{file=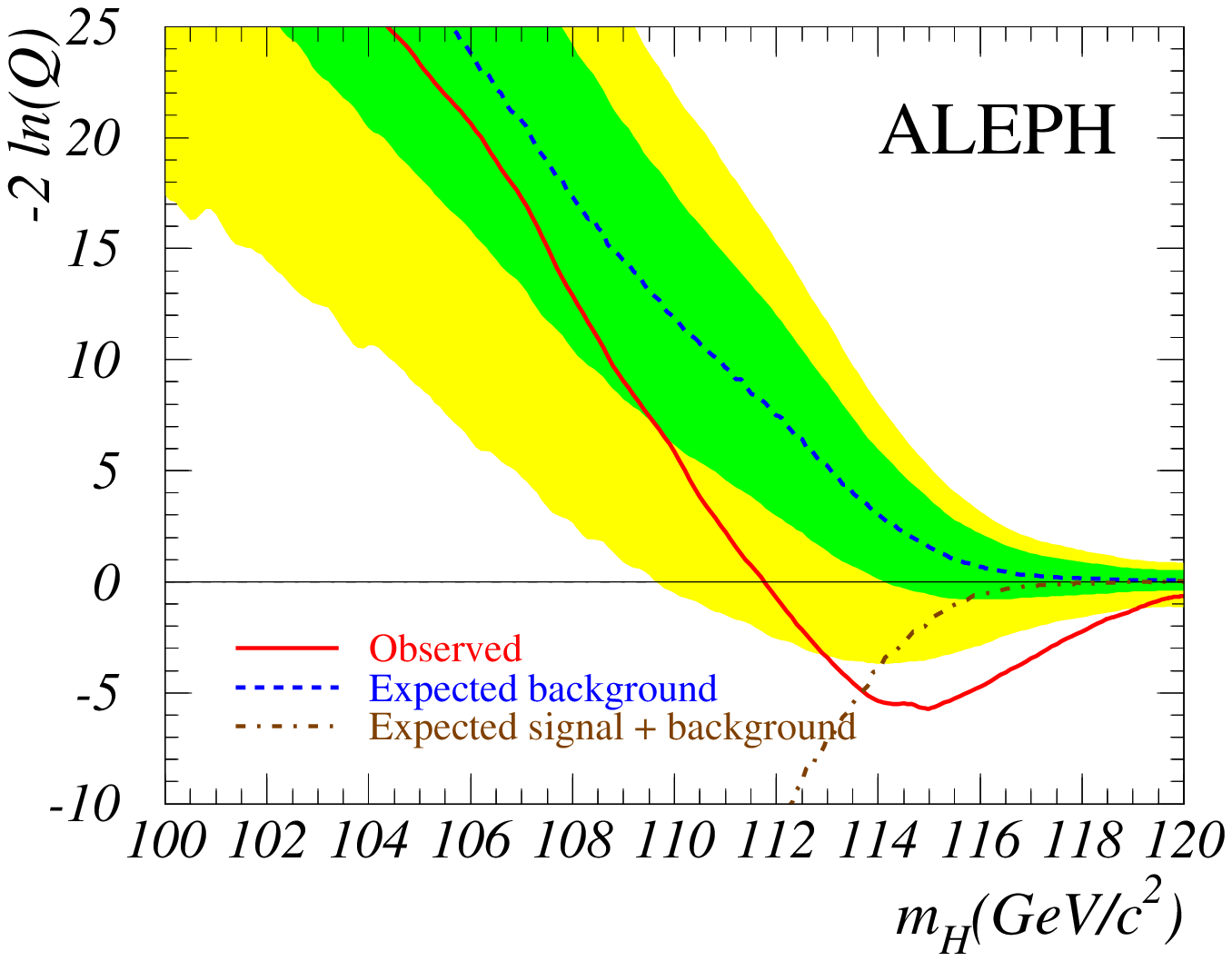,width=0.45\textwidth}
   \epsfig{file=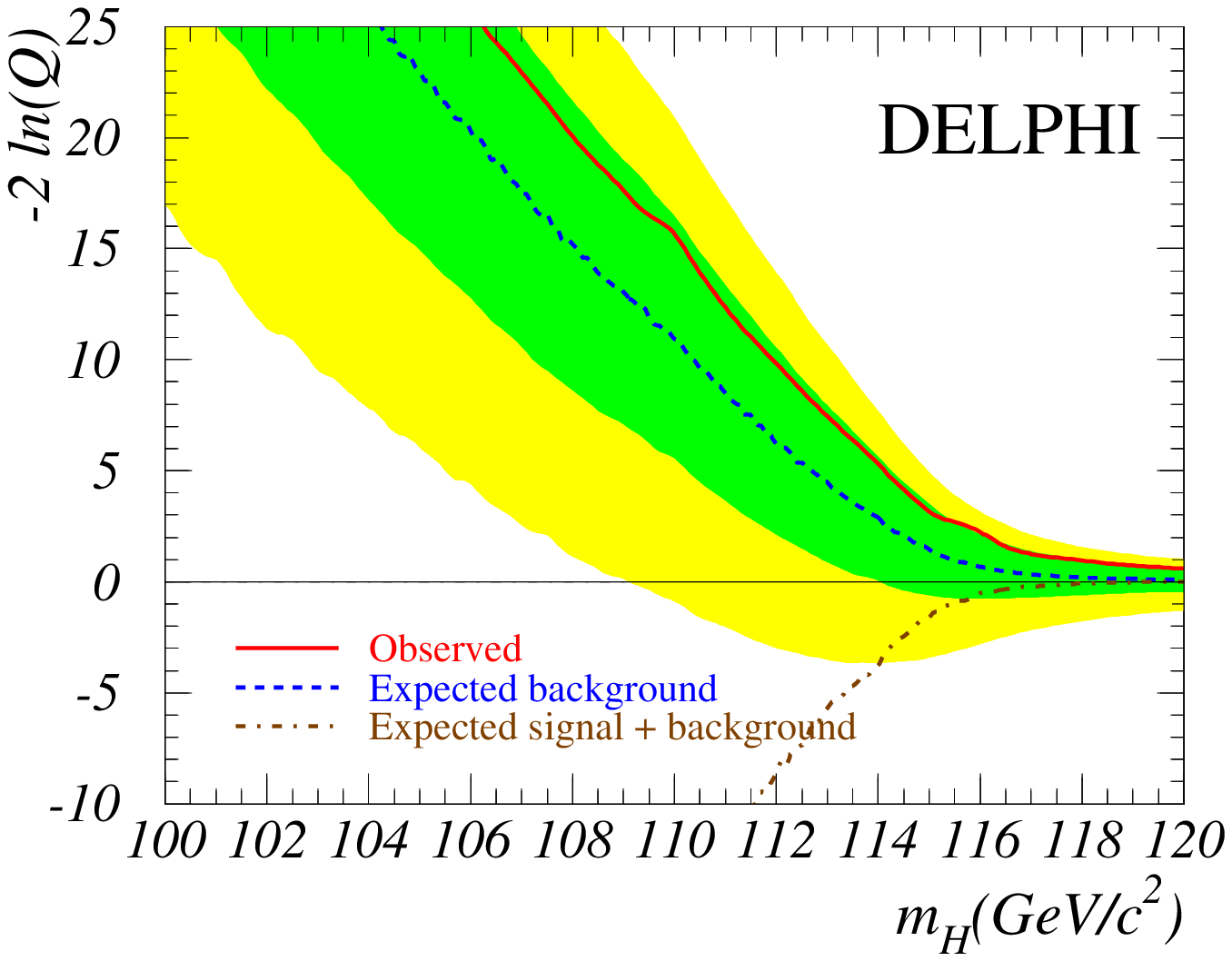,width=0.45\textwidth} \\
  \epsfig{file=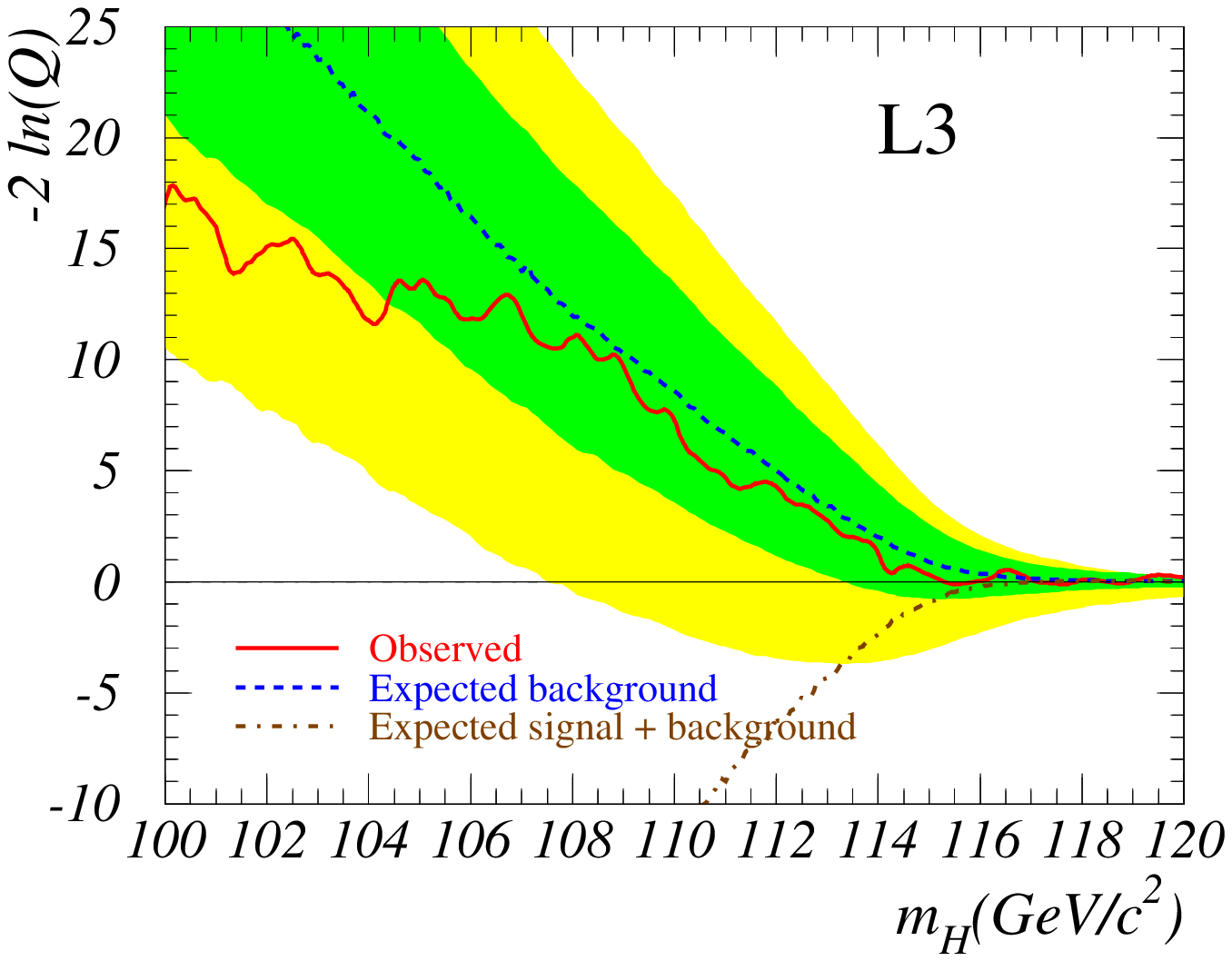,width=0.45\textwidth}
   \epsfig{file=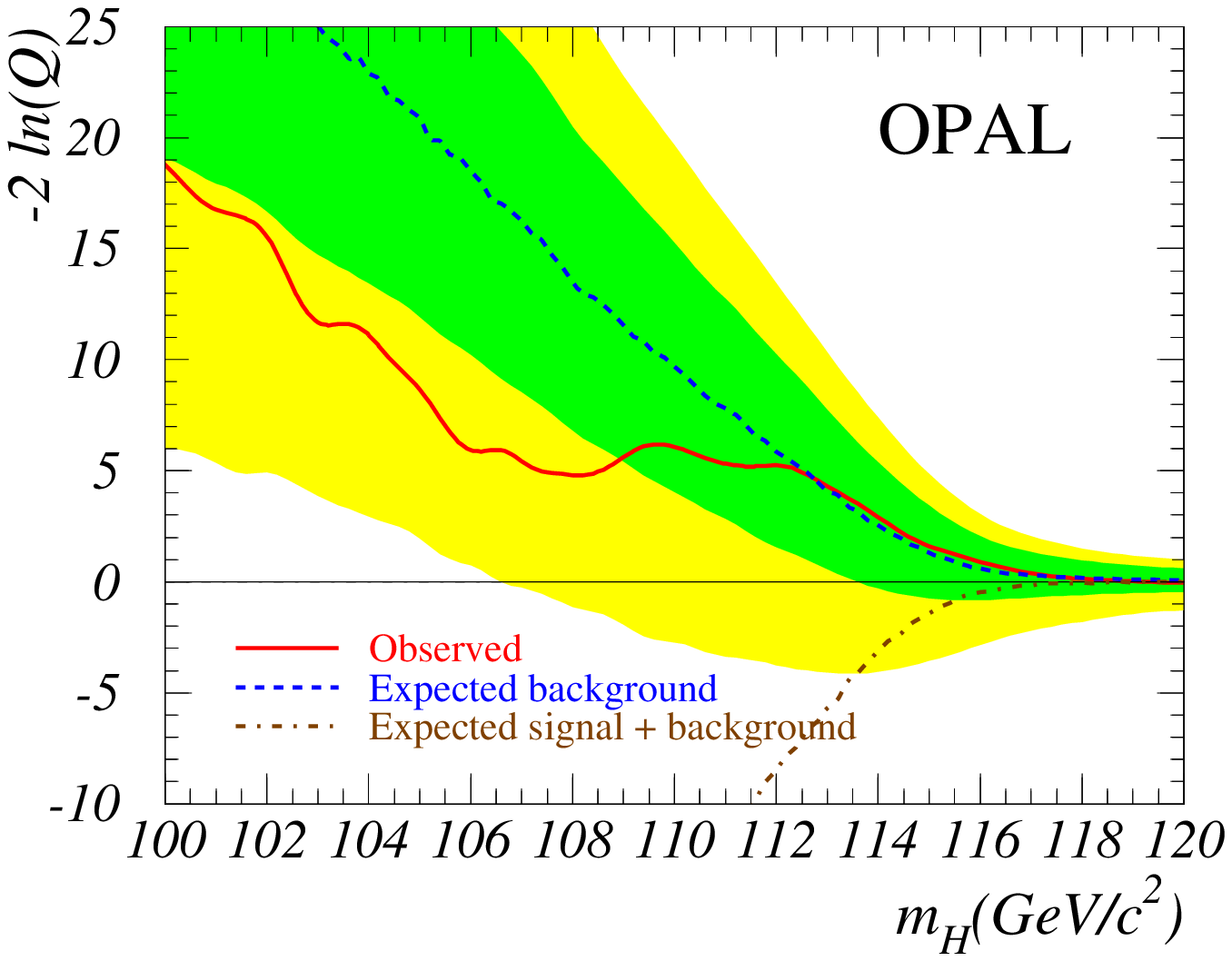,width=0.45\textwidth}

  \caption{\small Observed and expected behavior of the likelihood $-2\ln Q$ as a function of the test-mass
$m_H $ for the various experiments. The solid/red line represents
the observation; the dashed/dash-dotted lines show the median
background/signal+background expectations. The dark/green and
light/yellow shaded bands represent the 1 and 2 $\sigma$
probability bands about the median background expectation
\cite{bib-ADLO-ICHEP}. }\label{Fig-likelihood-adlo}
\end{figure}

\begin{figure}
  \centering
  \epsfig{file=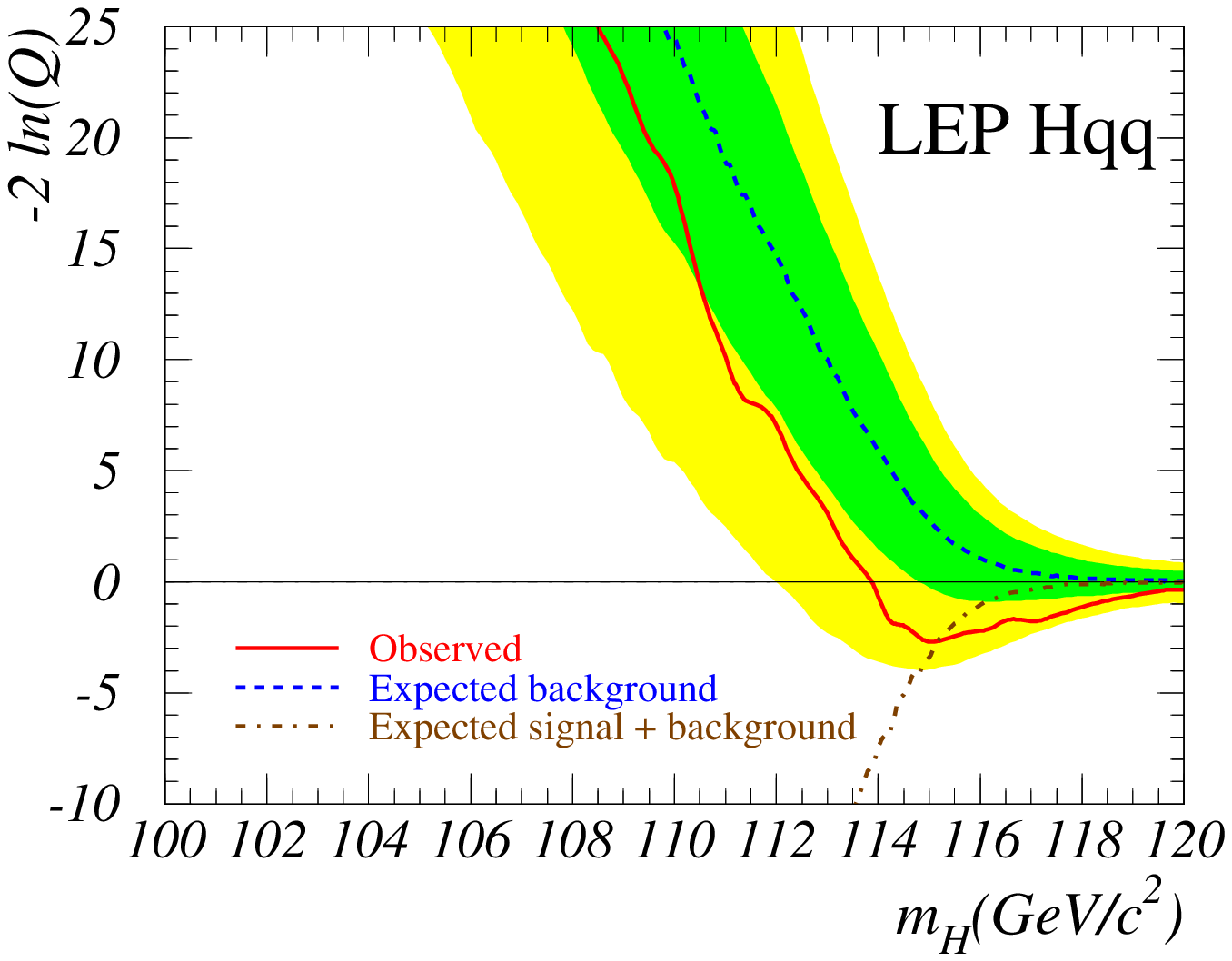,width=0.45\textwidth}
   \epsfig{file=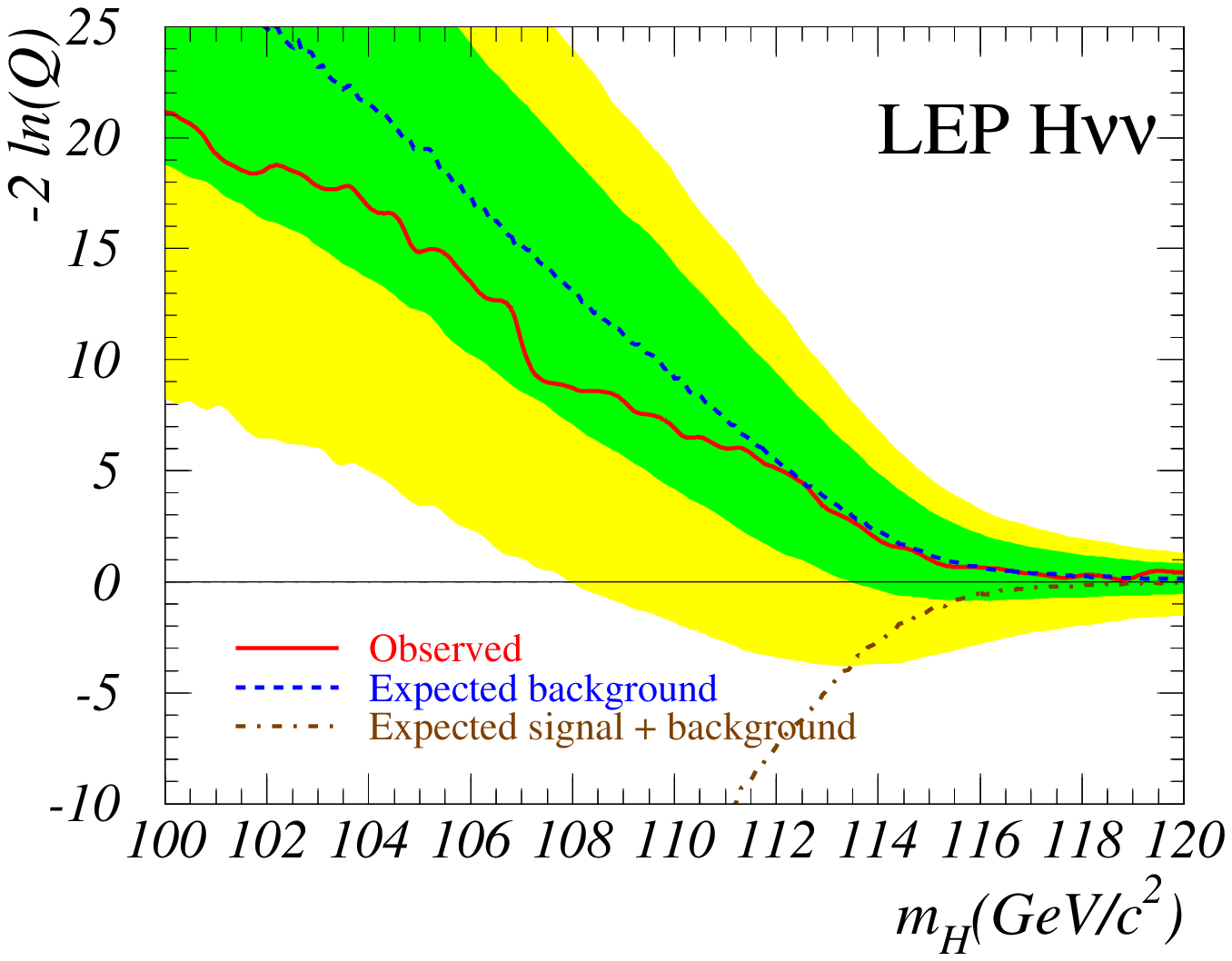,width=0.45\textwidth} \\
  \epsfig{file=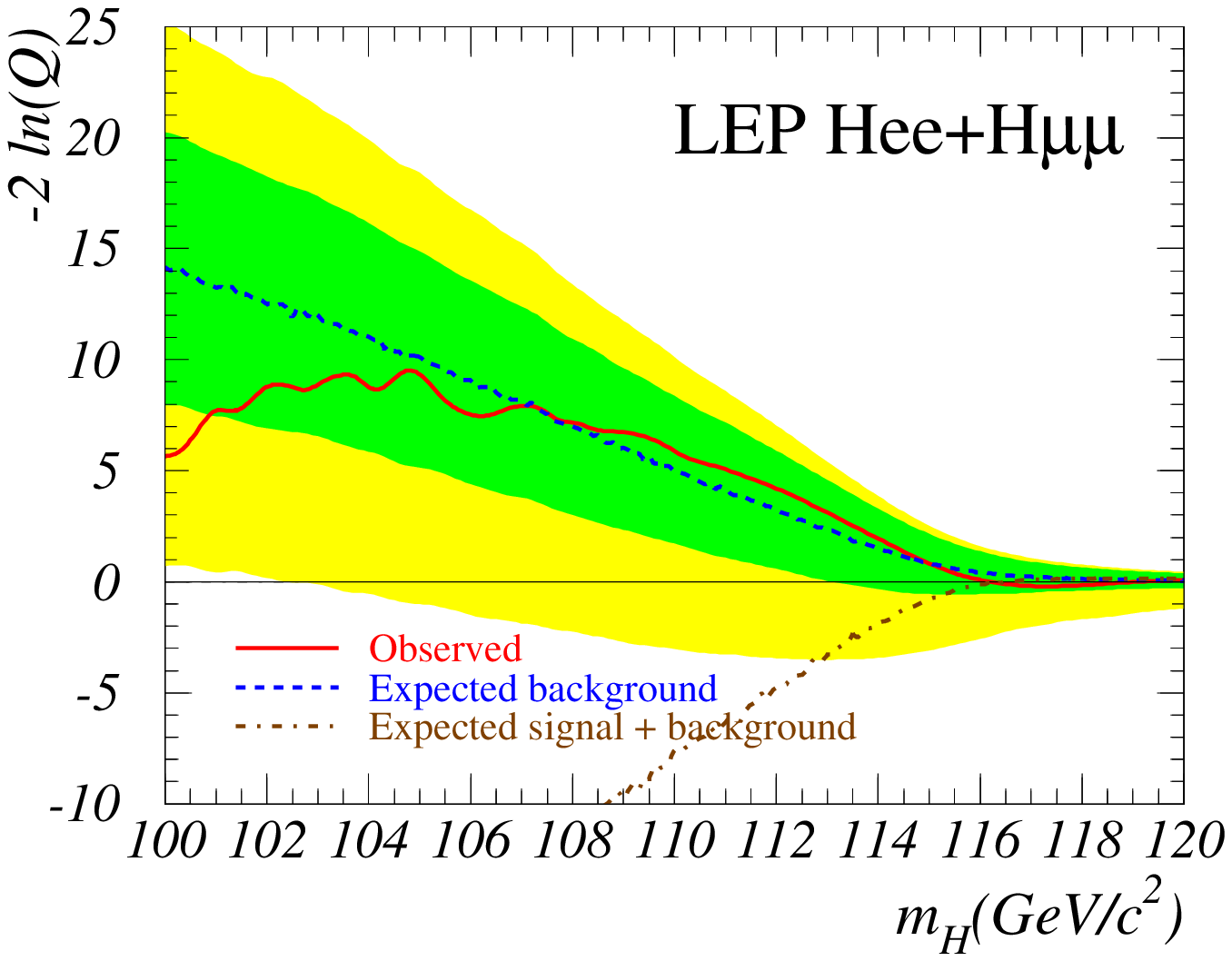,width=0.45\textwidth}
   \epsfig{file=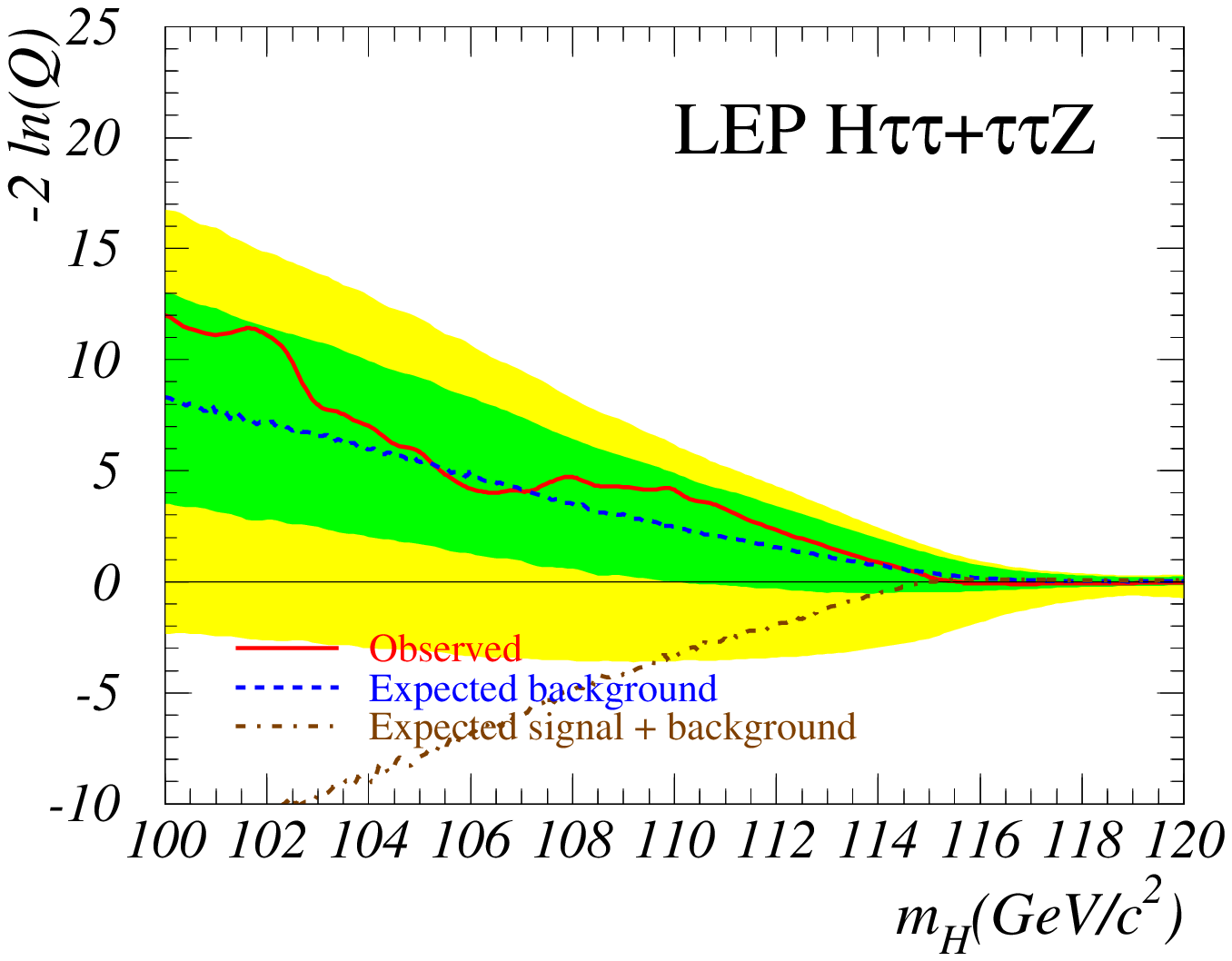,width=0.45\textwidth}

  \caption{\small Observed and expected behavior of the likelihood $-2\ln Q$ as a function of the test-mass
$m_H $ for the various Higgs search channels. The solid/red line
represents the observation; the dashed/dash-dotted lines show the
median background/signal+background expectations. The dark/green
and light/yellow shaded bands represent the 1 and 2 $\sigma$
probability bands about the median background expectation
\cite{bib-ADLO-ICHEP}. }\label{Fig-likelihood-channels}
\end{figure}

Now that we have learned how to read the likelihood plots let us
examine the LEP experimental outputs \cite{bib-ADLO-ICHEP}. The
likelihood outcome of the combined four LEP experiments is shown
in Figure \ref{Fig-likelihood-LEP}. This result can be sliced into
the different experiments (Figure \ref{Fig-likelihood-adlo}) or
the different search channels (Figure
\ref{Fig-likelihood-channels})\cite{bib-ADLO-ICHEP}.
 Shown are the \textit{b}-only median expectations with its 1 and 2$\sigma$ bands, the s+b expectation and the observed likelihood.
 The  broad minimum of the combined LEP likelihood from $m_H\sim 115-118$ GeV which crosses the
 expectation for \textit{s}+\textit{b} around $m_H\sim 116$ GeV can be interpreted as a
 preference for a Standard Model Higgs boson at this mass range, however, at
 less than the 2$\sigma$ level. When the LEP Higgs working group
 presented these results for the first time the significance was
 2.9$\sigma$ \cite{bib-pik}, and this relatively high significance
 generated a storm  which unfortunately turned out to be in a tea
 cup...

 The ALEPH observed likelihood has a 3$\sigma$ signal-like  behavior around
 $m_H\sim $ 114 GeV, which led the collaboration to claim a possible
 observation of a SM Higgs boson \cite{bib-aleph}. This behavior
 originated mainly from the 4-jet channel and
 its significance is reduced  when all experiments are combined.
 No other experiment or channel indicated a signal-like behavior.

 \section{Confidence Levels or How Probable is a Result?}

\begin{figure}
  \centering
  \epsfig{file=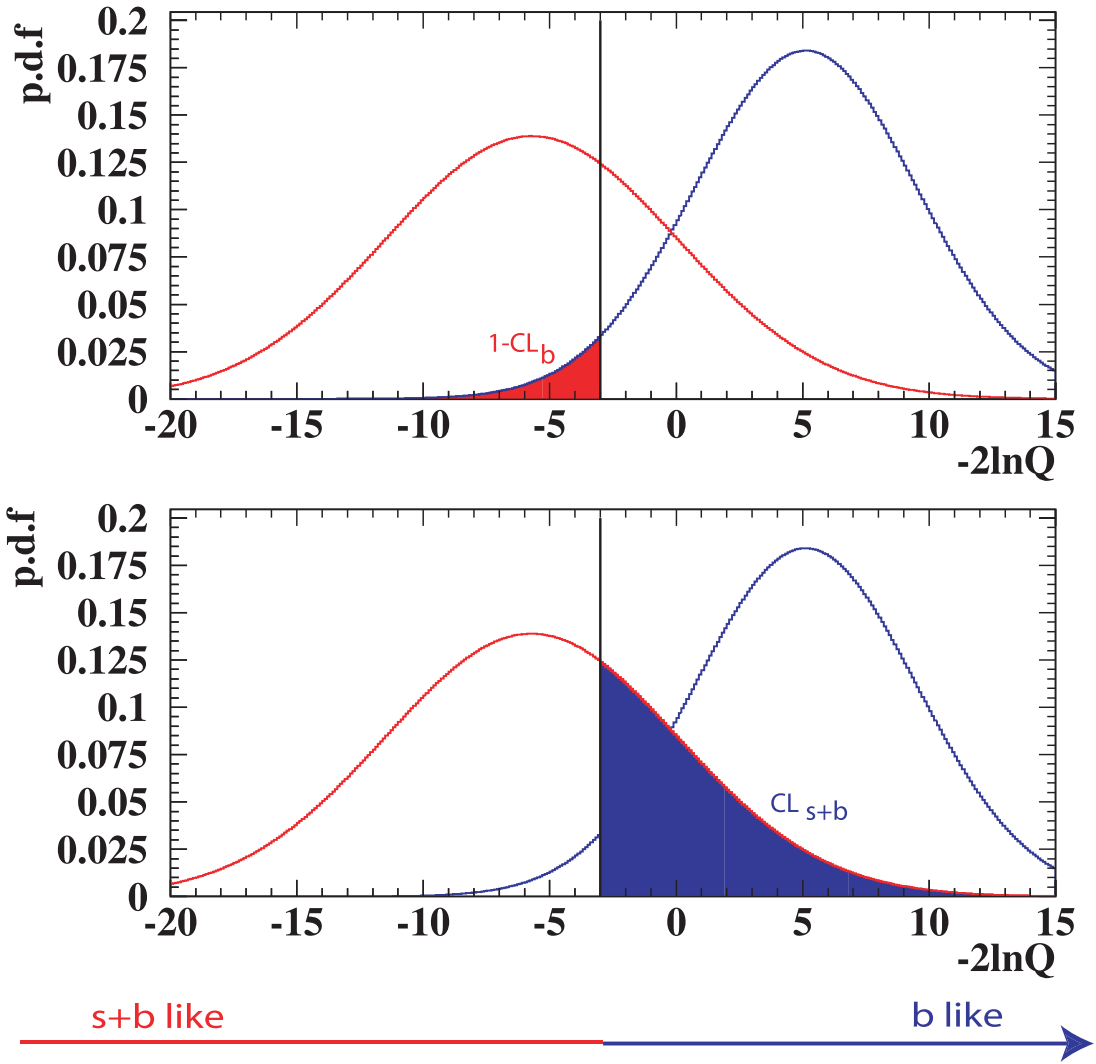,width=1.00\textwidth}

  \caption{An example of probability density functions (p.d.f's) for background only (solid blue) and signal+background (solid red) experiments.
  The red shaded
area, $1-CL_b$, measure the compatibility with the background
hypothesis while the blue shaded area, $CL_{s+b}$, the
compatibility with the signal+background hypothesis.
   Detailed explanations are given in the text.}\label{Fig-CL}
\end{figure}

\begin{figure}
  \centering
  \epsfig{file=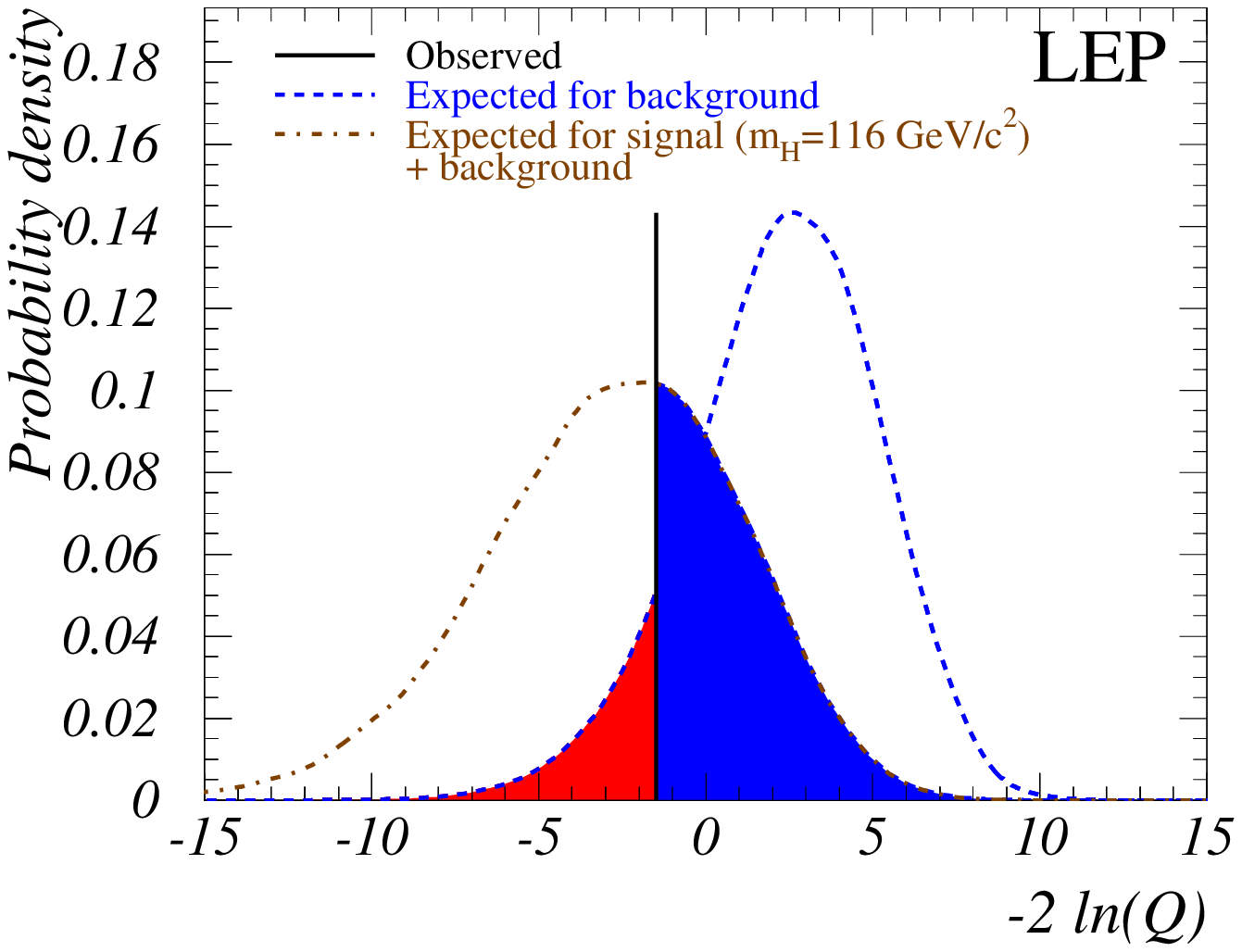,width=1.00\textwidth} \\
  \epsfig{file=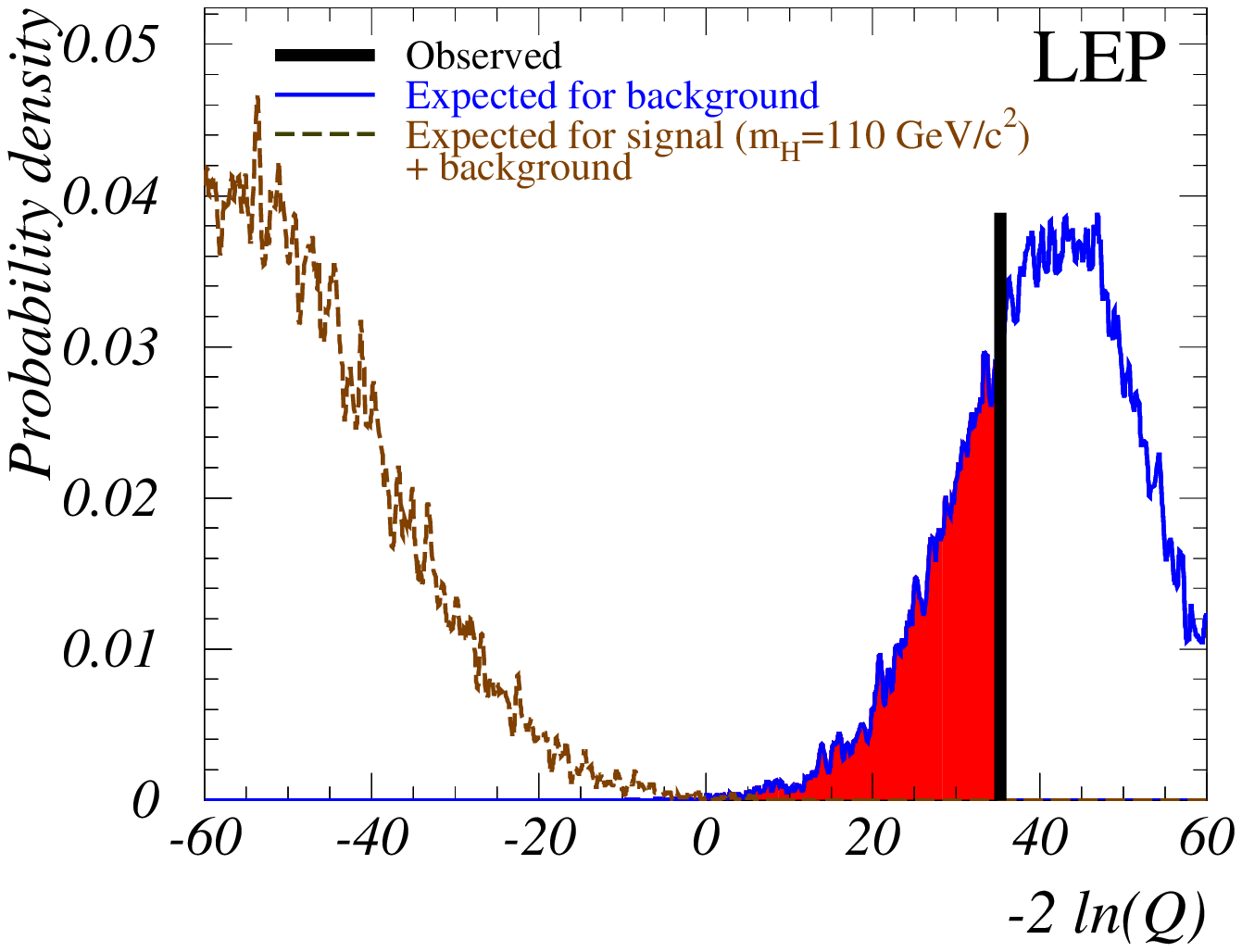,width=0.45\textwidth}
 \epsfig{file=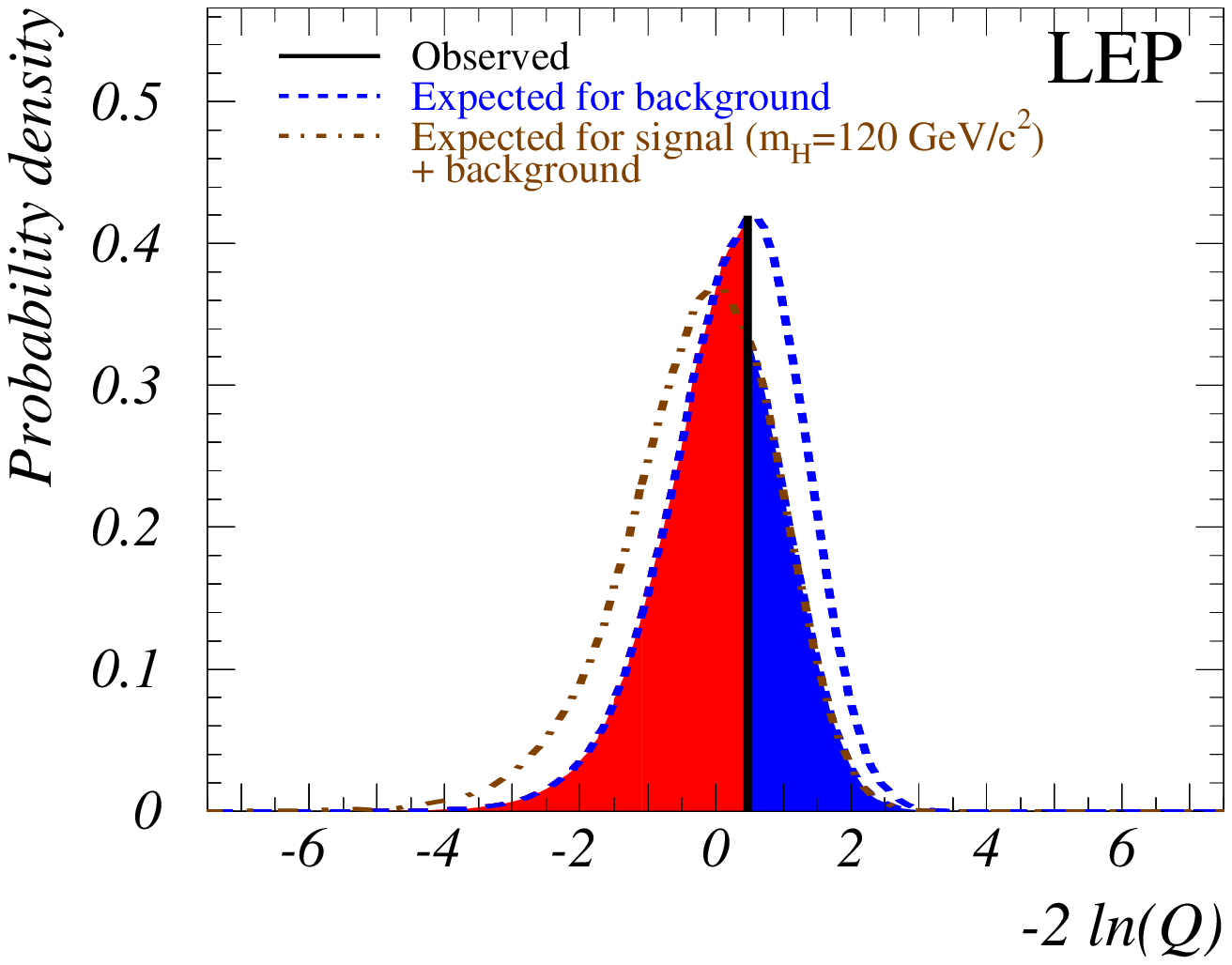,width=0.45\textwidth}

  \caption{\small Probability density functions corresponding to fixed test-masses, for the background
and signal+background hypotheses. The observed likelihood ratio
$-2\ln Q$ is indicated by the vertical line. The light/red shaded
areas, $1-CL_b$, measure the compatibility with the background
hypothesis and the dark/blue shaded areas, $CL_{s+b}$, the
compatibility with the signal+background hypothesis. Upper part:
test-mass $m_H =116$~GeV; lower part: $m_H =110$ (left) and
$120$~GeV (right) \cite{bib-ADLO-ICHEP}.}\label{Fig-pdfLEP}
\end{figure}

 Figure \ref{Fig-CL} shows the likelihood p.d.f. of our toy model with a test mass of $m_H \sim 115 $ GeV.
 Assuming an hypothetical observed likelihood of $-2\ln Q=-3$ the probability for a \textit{b}-only experiment to give a more
 \textit{s}+\textit{b}-like likelihood than the observed one is given by the area
 marked as $1-CL_b$, where $CL_b$ is the background confidence
 level. It is easy to see that the expectation value of this
 probability is 50\%, i.e. $\langle 1-CL_b \rangle=0.5$ irrespective of the test Higgs mass.
  One might
 also say that $CL_b$ measures the compatibility with the
 background hypothesis. The combined LEP p.d.f. is given in Figure
 \ref{Fig-pdfLEP}. For a test mass of $m_H=116$ GeV, the observed
 likelihood is such that $1-CL_b=0.099$ \cite{bib-ADLO-ICHEP}, i.e.,
 in 9.9\% of background-only experiments, we expect to observe a result at least as
 signal-like  as we observe.
 In the bottom plots of this Figure, the p.d.f's are shown for Higgs test masses of 110 and 120
 GeV where the observed probability is clearly consistent with the
 background hypothesis.
 These probabilities can be translated into a
 significance. A Gaussian approximation is used \cite{bib-pdg}.
 Table \ref{tab:significance}
 \begin{table}[htb]
\begin{center}
\begin{tabular}{|c||c|c|c|c|c|}
\hline
 $1-CL_b$  & $0.32$     &  $0.046$   &  $2.7\times 10^{-3}$ &  $6.3\times 10^{-5}$  &  $5.7\times 10^{-7}$
 \\ \hline
           &  1$\sigma$ &  2$\sigma$ &3$\sigma$             & 4$\sigma$             & 5$\sigma$             \\

\hline
\end{tabular}
\caption{\small Significances \label{tab:significance}}
\end{center}
\end{table}
shows the correspondence between
 $1-CL_b$ and the resulting significance. Therefore, the
 combined LEP observed likelihood corresponds to a significance
 below  2$\sigma$.

\begin{figure}
  \centering
  \epsfig{file=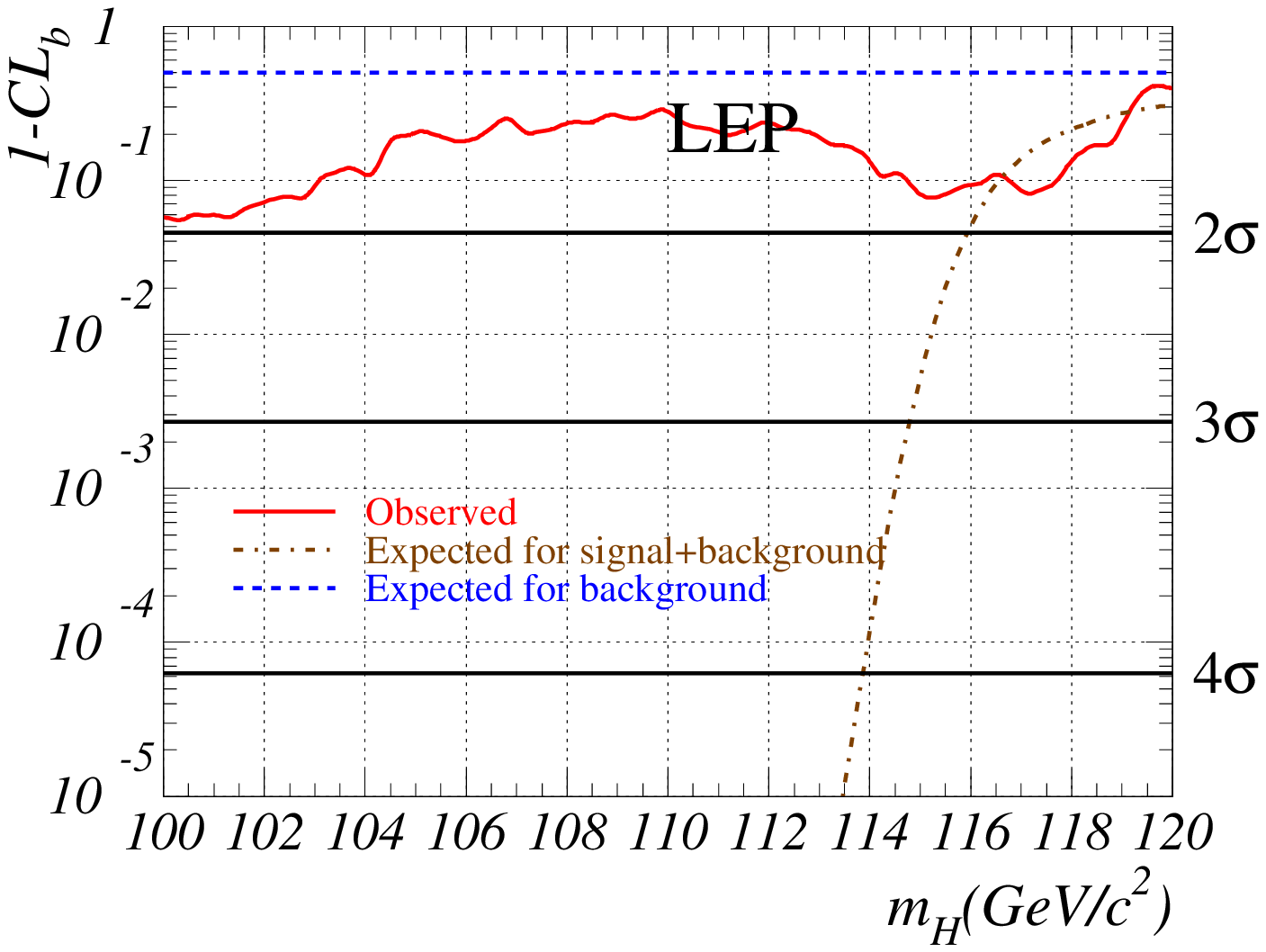,width=0.60\textwidth} \\
   \epsfig{file=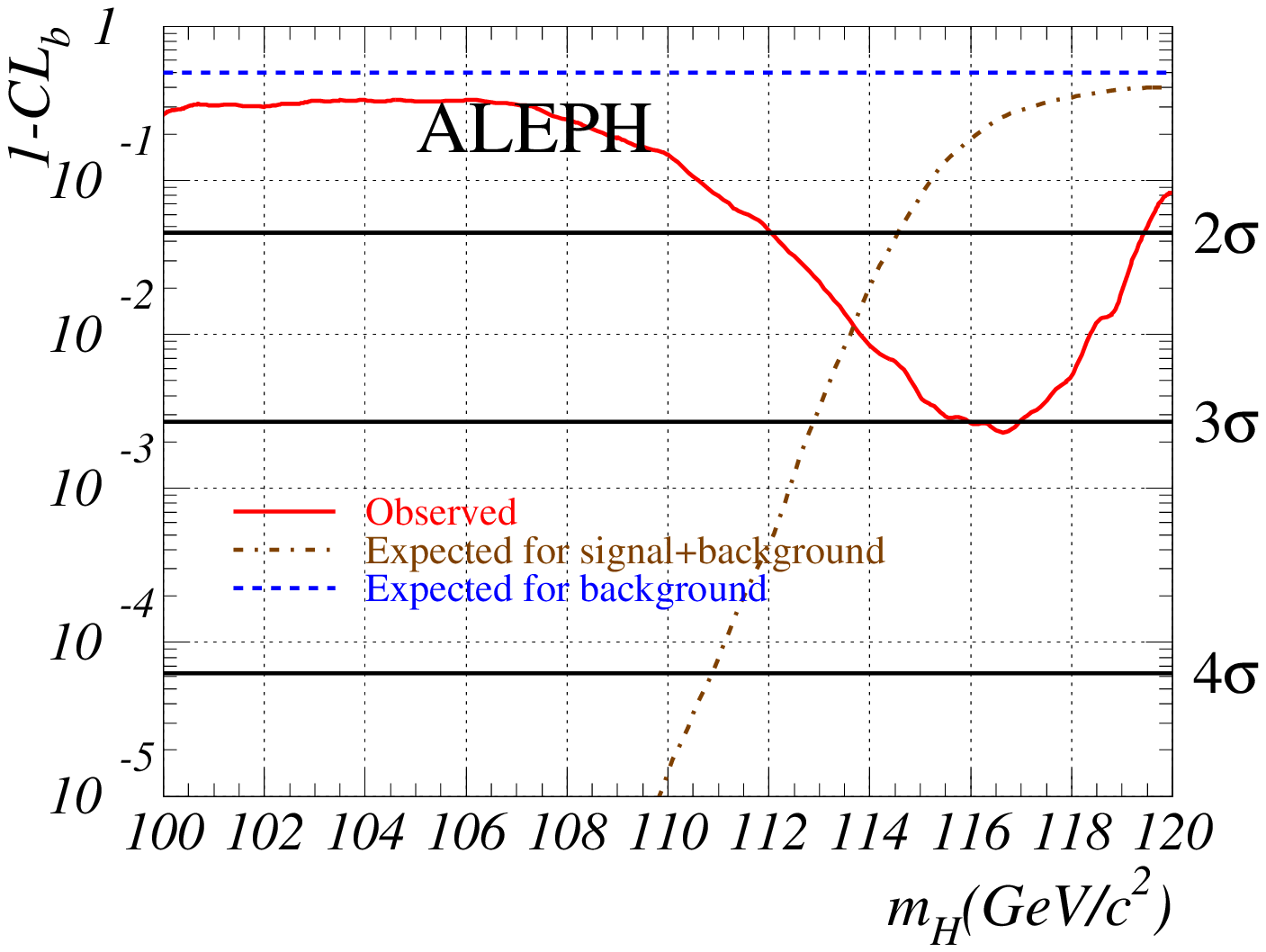,width=0.45\textwidth}
  \epsfig{file=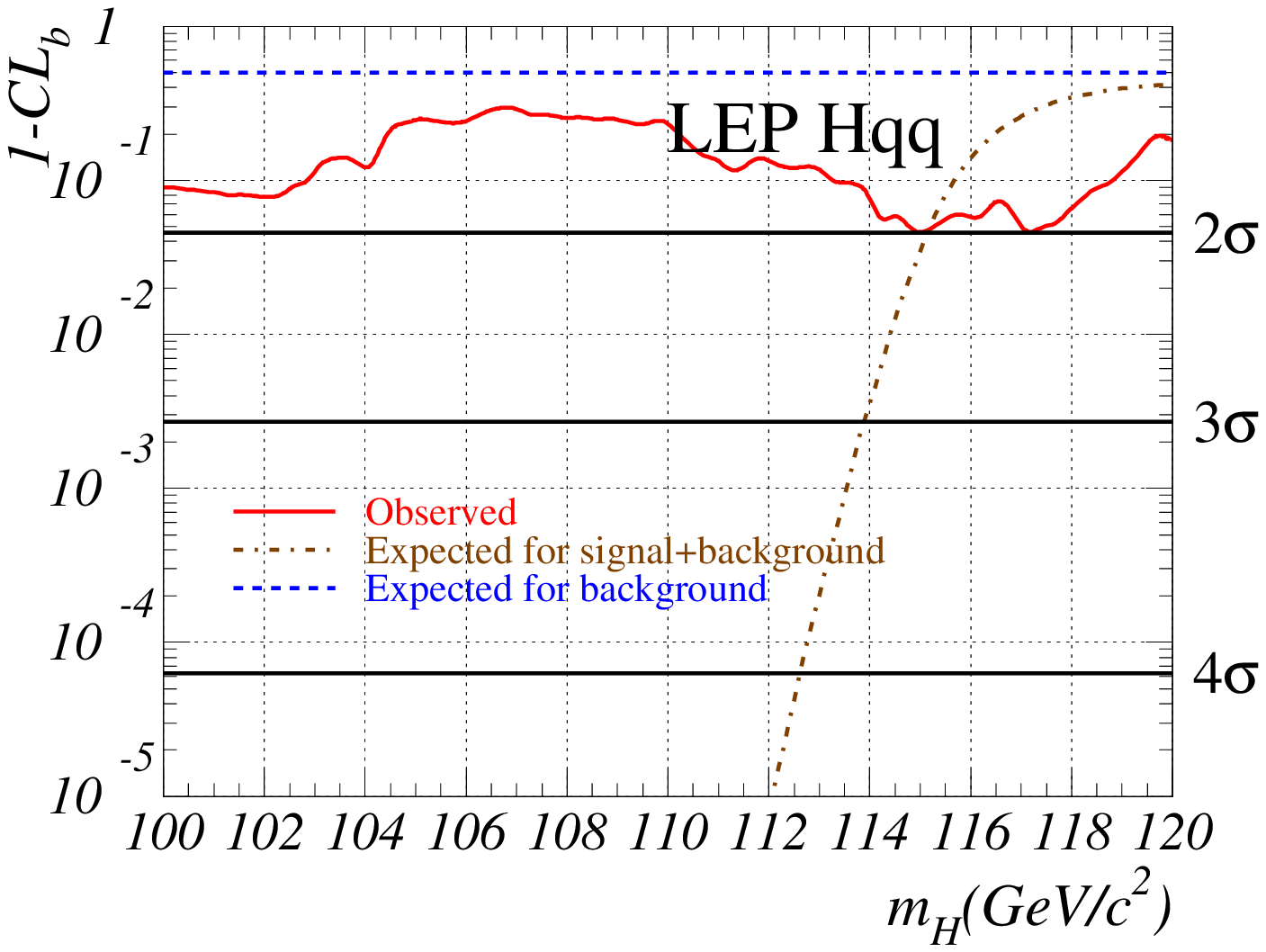,width=0.45\textwidth}

  \caption{\small Probability $1-CL_b$ as a function of the test-mass $m_H$. Solid/red line: observation;
dashed/dash-dotted lines: expected probability for the
background/signal+background hypotheses. The horizontal solid
lines indicate the levels for 2$\sigma$, 3$\sigma$ and 4$\sigma$
deviations from the background hypothesis. The top plot is the
combined LEP result while the bottom plots show the ALEPH and the
LEP 4-jet channel results\cite{bib-ADLO-ICHEP}. }\label{Fig-CLLEP}
\end{figure}

 Figure \ref{Fig-CLLEP} (top) shows the probability $1-CL_b$ as a function
 of the test mass $m_H$ \cite{bib-ADLO-ICHEP}. Shown are the median probability expected for \textit{b}-only
 (dashed) and \textit{s}+\textit{b} (dash-dotted) experiments, and the observed probability (solid
 red line). One can see that even though the observed probability
 is compatible with a $\sim$116 GeV Higgs boson, the sensitivity in
 this vicinity is less than  2$\sigma$, and LEP did not really
 have the sensitivity to observe a Higgs boson heavier than 115 GeV
 with more than 3$\sigma$ significance
 (this can be seen by the intersection of the dash-dotted line with the horizontal 3$\sigma$ line).
  In fact, The LEP sensitivity to observe a SM Higgs boson
at greater than 3$\sigma$ significance extends up to about 115 GeV
and up to about 116 GeV for observation at greater than 2$\sigma$.

  The bottom plots of Figure \ref{Fig-CLLEP} show the probability
  $1-CL_b$ where the ALEPH excess of candidate events in the 4-jet
  channel manifests itself as a 3$\sigma$ deviation for a 116
  GeV Higgs boson for ALEPH alone, and a  2$\sigma$ deviation in
  the combined LEP 4-jet channel.
  Note that ALEPH
  stand-alone sensitivity to observe a Higgs boson at the
  3$\sigma$ level extends up to $m_H\sim 113$ GeV, which is of course less than the combined LEP sensitivity.
  The ALEPH excess should therefore be interpreted as a fluctuation in that context.

  In case there are no clear indications for discovery,
one would like to interpret the search results in terms of
exclusion. The probability $CL_{s+b}$, shown as the blue area in
the bottom plot of Figure \ref{Fig-CL},  measures the
compatibility of the experiment with the \textit{s}+\textit{b}
hypothesis. There is no way to directly measure the signal
Confidence Level, $CL_s$ because of the presence of significant
background. A bigger $CL_{s+b}$ means the experimental result is
more \textit{s}+\textit{b}-like, but not necessarily more
\textit{s}-like due to the relative fluctuations of the
background. Therefore, if $CL_{s+b}$ is small, say, less than 5\%,
one can \emph{exclude} the \textit{s}+\textit{b} hypothesis at
more than 95\% Confidence Level, but that does not mean that the
signal hypothesis is excluded at that level. An example is given
in Table \ref{tab:clb} \cite{bib-ADLO-ICHEP}
\begin{table}[htb]
\begin{center}
\begin{tabular}{|l|cc|}
\hline
                 &      $1-CL_b$              &   $CL_{s+b}$        \\
\hline\hline
LEP              &         0.099              &    0.369         \\
\hline
ALEPH            &   $2.41\times 10^{-3}$     &    0.956         \\
DELPHI           &         0.874              &    0.033          \\
L3               &         0.348              &     0.408          \\
OPAL             &         0.543              &     0.208          \\
\hline
Four-jet         &   $5.70\times 10^{-2}$     &     0.676         \\
All but four-jet &         0.368              &     0.217         \\
\hline
\end{tabular}
\caption{ The background confidence level $1-CL_b$  and the
signal+background confidence level $CL_{s+b}$ at $m_H =116$~GeV,
for all LEP data combined and for various subsets of the data. The
numbers for the four-jet and all-but-four-jet final states are
obtained with the data of the four experiments combined. Table
taken from \cite{bib-ADLO-ICHEP}.} \label{tab:clb}
\end{center}
\end{table}
where the background confidence level $1-CL_b$  and the
signal+background confidence level $CL_{s+b}$ at $m_H =116$~GeV,
for all LEP data combined and for various subsets of the data are
shown . Note that the DELPHI $CL_{s+b}$ probability is 3.3\%. That
does not mean that DELPHI could exclude a 116 GeV Higgs boson
signal hypothesis, but rather that DELPHI can exclude a 116 GeV
\textit{s}+\textit{b} hypothesis. However, there is a
$1-CL_b=87.4\%$ probability for the DELPHI background to fluctuate
and give a signal-like observation. It was  to take this
probability into account that one \textit{apriori defined} the
signal Confidence level to be $CL_s=\frac{CL_{s+b}}{CL_b}$
\cite{bib-AlexRead}. This way the signal hypothesis for $m_h=116 $
GeV is excluded by definition at only the 73.8\% Confidence Level,
$CL=1-CL_s=1-0.033/0.126=0.738$. This procedure for constructing
the signal confidence level is obviously conservative since the
coverage probability is in general greater than the $CL$.

Of the three plots shown in Figure \ref{Fig-pdfLEP} it is clear
now that  a 110 GeV Higgs boson, where the $CL_{s+b}$ probability
is too small to be seen, can be excluded by LEP. This is also seen
in Table \ref{tab:clb} where none of the experiments alone or the
combined LEP can exclude a 116 GeV Higgs boson.

\begin{figure}
  \centering
  \epsfig{file=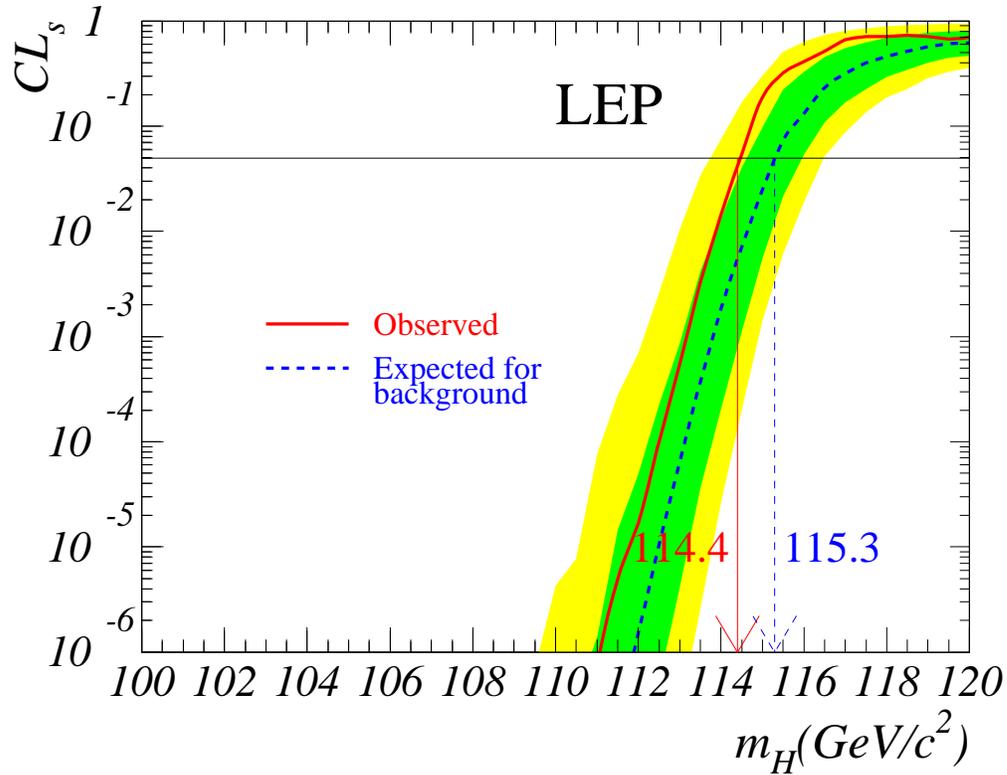,width=1.00\textwidth}

  \caption{\small Confidence level $CL_s$ for the signal+background hypothesis. Solid/red line: observation; dashed line:
median background expectation. The dark/green and light/yellow
shaded bands around the median expected line correspond to the 1
and 2 $\sigma$ probability bands computed with a large number of
simulated background experiments. The intersection of the
horizontal line at $CL_s = 0.05$ with the observed curve defines
the 95\% confidence level lower bound for the mass of the Standard
Model Higgs boson\cite{bib-ADLO-ICHEP}.}\label{Fig-exclusion}
\end{figure}

The exclusion power of LEP  is illustrated in Figure
\ref{Fig-exclusion} \cite{bib-ADLO-ICHEP} where the median
expected signal confidence level, $\langle CL_s^\rangle$, as a
function of the Higgs test mass together with its 1 and  2$\sigma$
bands is shown. Also shown is the observed $CL_s$ confidence
level. The intersection of the horizontal line at $CL_s=5\%$ with
the observed and median expected curves give the 95\% observed and
expected  exclusion \textit{CL}. Thus LEP excluded, at
 95\% \textit{CL}, a SM Higgs boson with a mass below 114.4
GeV, while it had the sensitivity to exclude a 115.3 GeV Higgs
boson. The small excess of data candidate events around 115--116
GeV is responsible for the slight reduction of the observed Higgs
lower mass limit with respect to its potential exclusion
sensitivity.

\section{Conclusion}

LEP Standard Model Higgs search results were reviewed with the
emphasis  on a pedagogical explanation of the statistical
procedure based on a toy model constructed specifically for this
purpose.

\section{Acknowledgements}
We would like to thank Moshe Kugler , Rick Van Kooten and Alex
Read for their useful suggestions and illuminations after a
critical reading of this note.
 One of the authors (E.G.) would like to
thank the following people: Peter Zerwas for inviting him to
summarize the LEP Higgs status in SUSY02 at DESY Hamburg, and
thereby initiating this work; Alex Read for teaching him
everything he knows about Higgs statistics, Chris Tully for giving
him a private LEP Higgs statistics lesson and Inbar Mor for the
good vibes and support while completing this work.

\newpage

\end{document}